\documentclass[lettersize,journal]{IEEEtran}
\usepackage{amsmath,amsfonts}
\usepackage{algorithmic}
\usepackage{algorithm}
\usepackage{array}
\usepackage[caption=false,font=normalsize,labelfont=sf,textfont=sf]{subfig}
\usepackage{textcomp}
\usepackage{stfloats}
\usepackage{url}
\usepackage{verbatim}
\usepackage{graphicx}
\usepackage[table]{xcolor}
\usepackage{cite}
\usepackage[T1, OT1]{fontenc}
\DeclareTextSymbolDefault{\dh}{T1}
\usepackage{balance}

\usepackage{subcaption}
\usepackage{tabularx}
\usepackage[dvipsnames]{xcolor}
\usepackage[numbers]{natbib}
\usepackage[normalem]{ulem}
\usepackage{enumitem}
\usepackage{pifont}
\usepackage{pgfplots}
\usepackage{array}
\usepackage{lipsum}
\usepackage[utf8]{inputenc}
\usepackage{hyperref}
\usepackage{tabularray}
\usepackage{graphicx}
\usepackage{caption}
\usepackage{multicol}
\usepackage{multirow}
\usepackage{ragged2e}
\usepackage[most]{tcolorbox}
\usetikzlibrary{patterns}
\UseTblrLibrary{booktabs} 
\usetikzlibrary{calc}
\hypersetup{
    colorlinks=true,
    linkcolor=blue,
    filecolor=magenta,      
    urlcolor=cyan,
    pdftitle={Overleaf Example},
    pdfpagemode=FullScreen,
    }
\usepackage{amsmath}
\definecolor{mygrayzero}{gray}{0.9}
\definecolor{mygrayone}{gray}{0.8}
\definecolor{mygraytwo}{gray}{0.7}
\definecolor{mygraythree}{gray}{0.6}
\definecolor{Mycolor}{HTML}{166666}
\definecolor{revision}{HTML}{D10000}
{\newcommand{\nbc}[3]{\colorbox{#3}{\bfseries\sffamily\scriptsize\textcolor{white}{#1}}
{\textcolor{#3}{\sf\small$\blacktriangleright$\textit{#2}$\blacktriangleleft$}}}}
{\newcommand{\nbc}[3]{}}
\definecolor{Mycolor}{HTML}{166666}

\newcommand\functionalmark{\raisebox{-0.2em}{\includegraphics[width=1em]{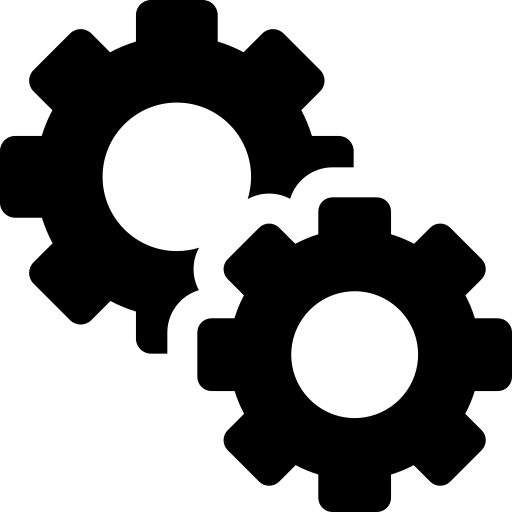}}} 
\newcommand\evolvemark{\raisebox{-0.2em}{\includegraphics[width=0.9em]{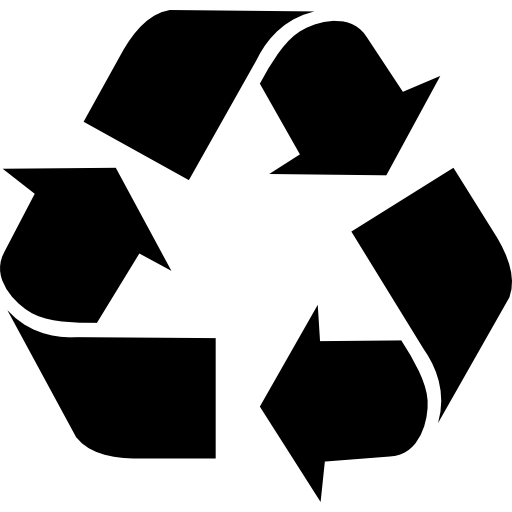}}} 
\newcommand\discussmark{\raisebox{-0.2em}{\includegraphics[width=1em]{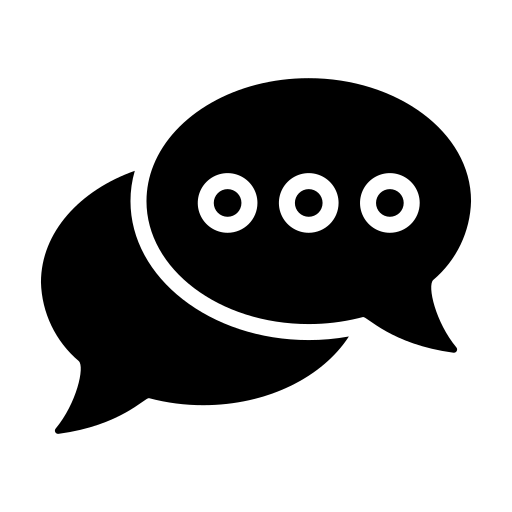}}}

\begin{document}

\title{“Go Home Copilot, You’re Drunk”: Understanding Developer Responses to Agent-Generated Code Review Comments}

\author{Shamse Tasnim Cynthia, Ratnadira Widyasari, Banani Roy, Ting Zhang and David Lo 

\thanks{Shamse Tasnim Cynthia and Banani Roy are with Department of Computer Science, University of Saskatchewan, Saskatoon, Canada}
\thanks{Ratnadira Widyasari and David Lo are with the School of Computing and Information Systems, Singapore Management University, Singapore}
\thanks{Ting Zhang is with the Department of Software Systems and Cybersecurity, Monash University, Melbourne, Australia}
}


\markboth{Journal of \LaTeX\ Class Files,~Vol.~14, No.~8, August~2021}%
{Shell \MakeLowercase{\textit{et al.}}: A Sample Article Using IEEEtran.cls for IEEE Journals}


\maketitle

\begin{abstract}
    Code review is a critical quality assurance practice in software engineering development, and AI coding agents are increasingly generating review comments on pull requests. However, little is known about how developers actually respond to such agent-generated feedback. 
    In this paper, we present the first large-scale empirical study on the resolution of agent-generated code review comments. We analyze $54{,}791$ comments generated by five widely used coding agents (i.e., Copilot, Cursor, Codex, Devin, and Claude) across $342$ Python repositories on GitHub. We examine (1) resolution rates across agents and comment types, (2) the role of developer experience, and (3) characteristics that influence comment usefulness. 
    Our results show that resolution rate varies considerably across agents, with Copilot accounting for the majority of resolved comments (72.9\%). Core developers resolve the majority of agent-generated feedback, particularly for \textit{design} and \textit{evolvability}-related comments, while peripheral developers are more involved in resolving \textit{functional defect} comments. Through open card sorting of 470 unresolved comment discussions, we identify \textit{ten} discussion patterns explaining why comments remain unresolved, with \textit{incorrect suggestions} and \textit{intentional design decisions} being the most prevalent. Finally, our analysis reveals that the presence of an inline \textit{code suggestion} is the strongest predictor of comment resolution, while lengthy and complex comments are less likely to be acted upon. Our findings provide insights for improving AI-generated code review feedback and its integration into development workflows.  
\end{abstract}


\section{Introduction} \label{sec:introduction}
Modern software development increasingly relies on code review as a central quality gate for preventing defects before integration, with developers spending 10-15\% of their time on this task~\cite{thongtanunam2022autotransform, goldman2025types}. While code review ensures removing functional defects, improving design quality, maintainability, and reliability~\cite{mcintosh2016empirical, morales2015code, sadowski2018modern}, it is a human-intensive process where developers have to manually review and revise code~\cite{thongtanunam2022autotransform}. Large-scale industrial systems illustrate this challenge. For example, Microsoft Bing processes roughly 3K code reviews per month~\cite{rigby2013convergent}, while at Google, developers spend on average about 60 minutes actively shepherding a change from submission for review to final integration~\cite{frommgen2024resolving}. To cope with the growing review workloads, organizations are beginning to deploy AI-powered code review agents that automatically generate review comments, highlight potential issues, and propose changes~\cite{goldman2025types}. Although early studies indicate that these agents can surface useful findings and reduce reviewer effort, a substantial portion of their comments still requires developer validation and resolution. This raises concerns around trust, false positives, response latency, and potential integration friction ~\cite{goldman2025types, adhalsteinsson2025rethinking}. 

Recent studies have begun to investigate the effectiveness and characteristics of AI-generated code review comments in both industrial and open-source development environments. For example, Goldman et al. \cite{goldman2025types} shows that AI and human reviewers tend to emphasize different categories of issues, and that approximately 60–70\% of LLM-generated comments remain unresolved. In another study, Fr{\"o}mmgen et al. \cite{frommgen2024resolving} introduced an ML-assisted feature at Google to help developers resolve review comments more efficiently, reporting that 7.5\% of all code review comments were addressed using ML-suggested edits. Similarly, Sun et al. \cite{sun2025does} found that although adoption of such tools is increasing, their effectiveness varies considerably. In particular, comments that are concise, include code snippets, and are manually triggered are more likely to lead to code modifications. 

Despite recent advances, how developers respond to AI-generated code review comments remains underexplored. First, we lack a clear understanding of how resolution behavior varies across different agents. Second, prior work provides limited insight into how developer experience influences the acceptance of AI-generated feedback. Third, little is known about the discussion patterns that emerge when such comments remain unresolved. Understanding these aspects is essential to assessing how AI review agents integrate into collaborative workflows and what makes their feedback useful.

To address these gaps, we conducted an empirical study of $54{,}713$ agent-generated code review comments produced by three widely used agents: Copilot, Cursor, and Codex, from $341$ Python repositories on GitHub.
Our objectives are (1) to examine how frequently developers address review comments generated by AI agents and how this behavior differs across agent tools and types of comments, (2) to investigate how developer experience shapes the acceptance of agent-generated feedback and to characterize the discussions that occur when such comments remain unresolved as these discussions reveal limitations of agent feedback and barriers to its adoption, and (3) to identify features of agent-generated review comments that influence their usefulness to developers. We aim to provide insights into how agent-generated code review comments can be effectively integrated into modern code review practices.

%
We construct a large-scale annotated dataset of agent-generated comments and analyze it using a combination of automated and manual techniques, as such real-world datasets remain limited. We classify comments into 15 predefined categories and group contributors into core and peripheral developers to compare resolution behavior. We then apply open card sorting to identify themes underlying non-resolution, resulting in a taxonomy of 10 categories and 13 sub-categories. Finally, we analyze factors influencing comment usefulness, including structural features, quality metrics, and explanation types.
The contributions of this paper are: 
(1) Conducting a large-scale empirical study of $54{,}713$ agent-generated code review comments from three coding agents across $341$ GitHub repositories.
(2) Analyzing resolution behavior across agents and comment categories to understand adoption patterns. (3) Investigating the role of developer experience and identifying \textit{ten} categories of discussion patterns in unresolved comments. (4) Examining key characteristics of agent-generated comments across \textit{five} factors. 


\section{Motivating Example} \label{sec:methodology}
\begin{figure}[t]
    \centering
    \subfloat[Example 1.]{\includegraphics[width=\linewidth]{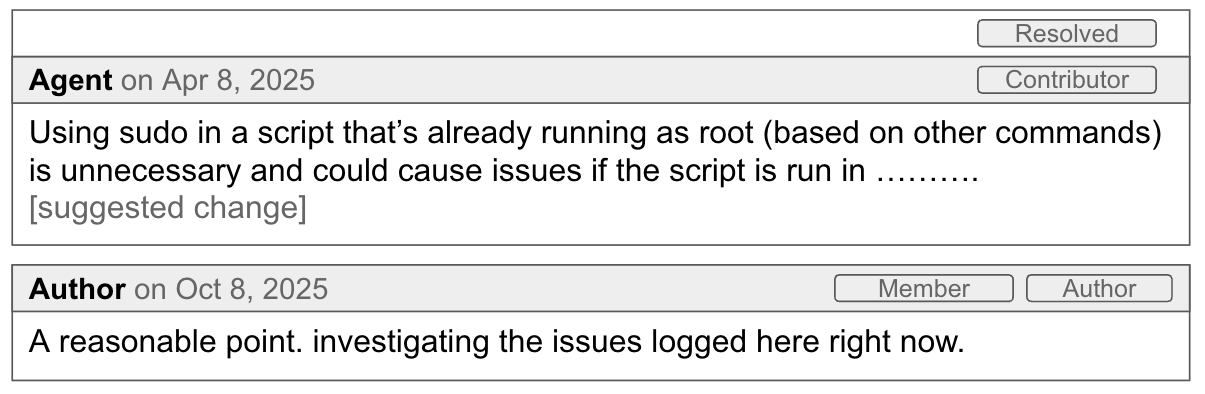}\label{fig:example1}}
    \hfill
    \subfloat[Example 2.]{\includegraphics[width=\linewidth]{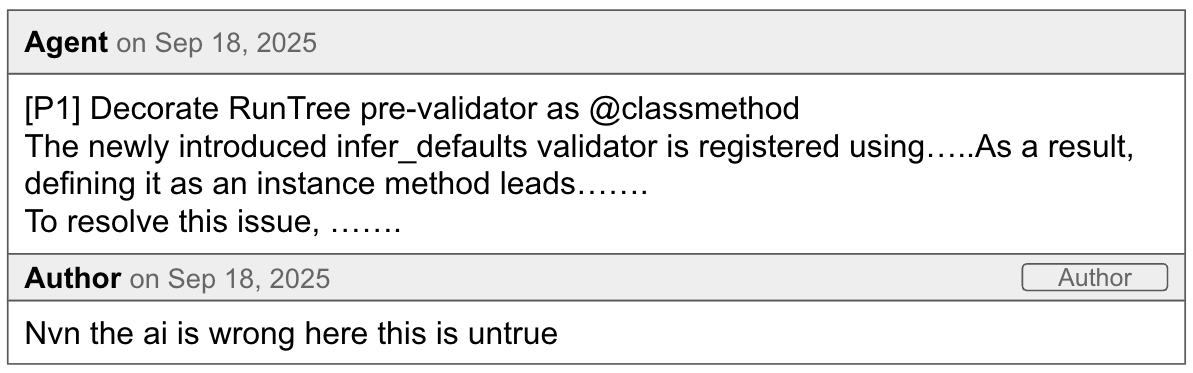}\label{fig:example2}}
    \caption{Examples of agent-generated review comments.}
    \vspace{-0.5cm}
    \label{fig:motivating-examples}
\end{figure}
Despite the growing adoption of AI coding agents in code review, their feedback is not always perceived as useful. Fig.~\ref{fig:motivating-examples} shows two contrasting examples. In Fig.~\ref{fig:example1}, the agent flags an unnecessary use of \texttt{sudo} and provides a rationale, which the developer acknowledges and acts upon \cite{compiler_explorer_pr_1740}. In Fig.~\ref{fig:example2}, the agent suggests a change with a detailed explanation, but the developer rejects it as incorrect \cite{langsmith_sdk_pr_2030}.
These examples highlight that usefulness depends not only on explanation but also on contextual accuracy and actionability. Incorrect or misaligned suggestions can undermine trust, motivating our investigation into the current state of agent-generated comments, how developers respond to them, and which characteristics make them effective.

\section{Methodology} \label{sec:methodology}
\begin{figure}
    \centering
    \includegraphics[width=\linewidth]{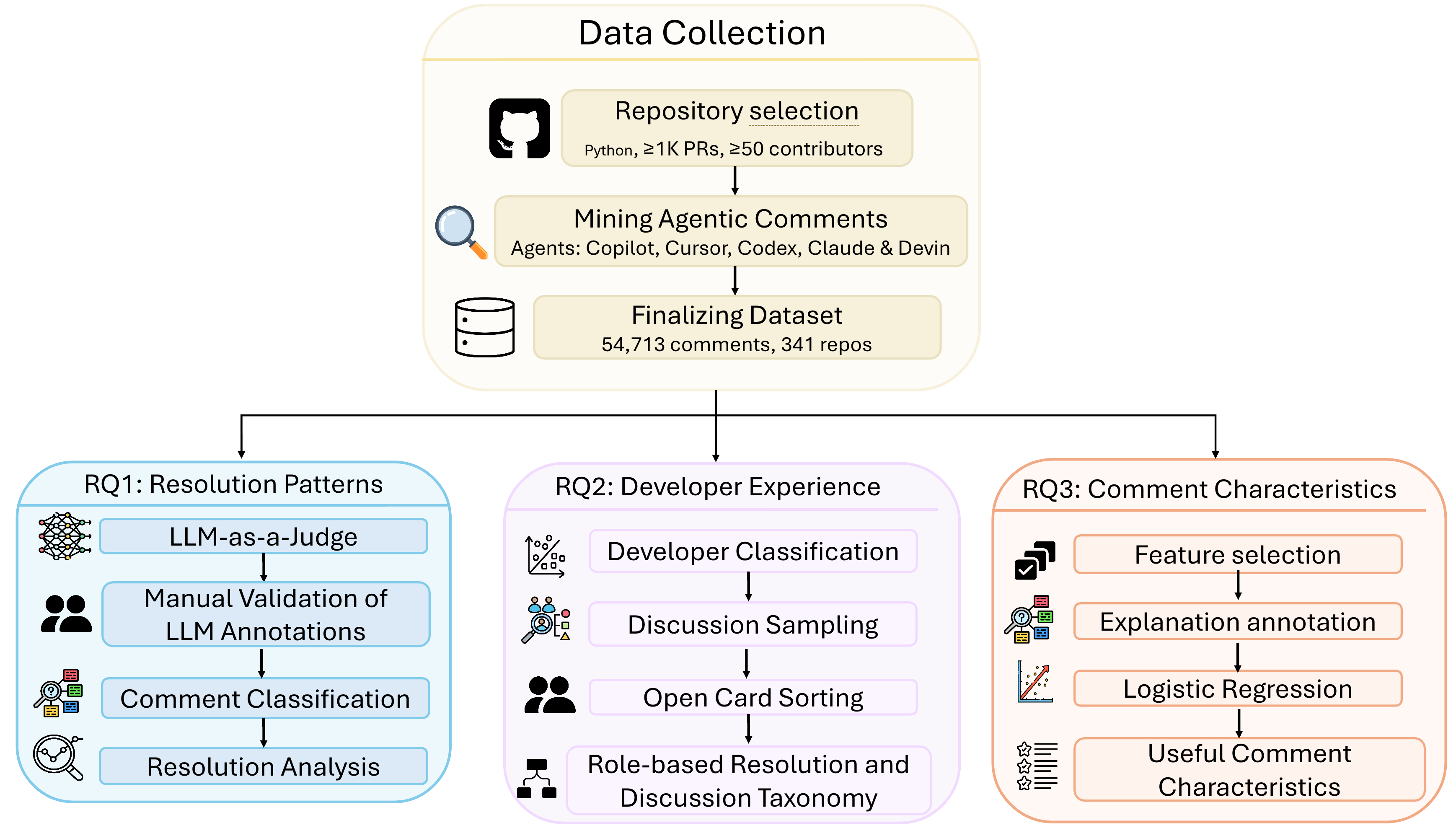}
    \caption{Overview of our approach}
    \label{fig:methodology}
    \vspace{-0.5cm}
\end{figure}
We aim to understand developer responses to agent-generated code review comments by addressing the following research questions:
\begin{itemize}[leftmargin=*] 
 \renewcommand\labelitemi{\ding{43}} 
    \item \textbf{RQ\textsubscript{1}. What proportion of agent-generated code review comments are resolved by developers across different agents, and how does resolution likelihood vary by comment type?}
    AI coding agents are increasingly participating in code review by automatically generating feedback on PRs. However, the practical usefulness of such comments depends on how developers act on them. To understand the effectiveness of agent-generated comments, we examine how frequently developers resolve these comments and whether resolution patterns differ across agents and comment categories.
    \item \textbf{RQ\textsubscript{2}. How does developer experience influence the acceptance of agent-generated review comments, and what discussion patterns emerge when comments remain unresolved?}
    Code review is inherently a socio-technical process where developer roles and experience can influence how feedback is interpreted and acted upon \cite{bosu2014impact}. We therefore investigate whether developers' experience differs in responding to agent-generated comments and analyze the discussion pattern that arises when such comments remain unresolved.
    \item \textbf{RQ\textsubscript{3}. What characteristics of agent-generated review comments make them useful to developers?} Not all agent-generated review comments provide equal value to developers \cite{goldman2025types}. Understanding what characteristics make agent-generated comments useful can inform the design of more effective code review agents. 
\end{itemize}
%
Fig.~\ref{fig:methodology} provides an overview of our methodology. We first collect a dataset of GitHub open-source projects and use it to address our RQs. For RQ1, we analyze resolution rates across agents and comment categories. For RQ2, we examine developer experience by comparing core and peripheral contributors and analyzing unresolved comments through open card sorting. For RQ3, we investigate how comment characteristics relate to usefulness using statistical analysis and regression modeling.

\textbf{Data Collection} 
We selected open-source projects with a code review process active in GitHub, i.e., frequent pull requests (PRs). Our focus is on review comments generated by widely used coding agents, initially targeting five popular tools: Copilot, Cursor, Codex, Devin, and Claude \cite{li2025rise}. Those projects must be active software systems, e.g., applications, libraries, frameworks, with many contributors.
%
We used SEART~\cite{dabic2021sampling}, a web-based platform for querying GitHub repositories, to identify candidate projects. 
It is widely used in prior studies and supports flexible selection criteria~\cite{bouraffa2025not, tufano2022using, decan2022use, da2025understanding}. We apply the following criteria: (i) Python projects (since Python is the predominant language for agentic workflow \cite{hasan2026empirical, openja2024empirical}), 
(ii) at least 1000 stars, (iii) at least 1000 PRs, (iv) at least fifty contributors, and (v) not a forked repository. The contributor threshold ensures an active code review process while filtering out personal projects~\cite{kalliamvakou2014promises}. SEART returned $1{,}246$ projects.

Next, we retained the projects with at least one PR created after December 2024, when the newest agent (Devin) was introduced, ensuring all studied agents could appear~\cite{li2025rise}, resulting in $1,120$ projects.
%
We identified repositories containing agent-generated code review comments using 
GitHub GraphQL~\cite{github_graphql} and REST APIs~\cite{github_rest}. First, we retrieved PR review threads and associated comments using the GraphQL API. 
Next, we used the REST API to obtain review metadata and identify reviewers. 
A comment was classified as agent-generated if it was associated with a PR reviewed by a known AI coding agent, determined by matching reviewer \textit{login} names against predefined patterns, i.e., \texttt{copilot-pull-request-reviewer}, \texttt{claude}, \texttt{cursor}, \texttt{chatgpt-codex-connector}, and \texttt{devin-ai-integration} \cite{robbes2026promises, li2025rise, chatgpt_codex_connector}. A repository was retained if it contained at least one such instance, resulting in a final dataset of $342$ repositories.
While this approach enables scalable identification of agent-generated comments, it relies on naming conventions and may miss agents with non-standard identifiers. 
To mitigate this, we manually inspected a subset of repositories and confirmed that the detected accounts correspond to automated review agents.

To ensure sufficient time for review completion and agent interaction, we collected data in early 2026 but restricted our analysis to PRs created up to December 2023, allowing a one-year buffer for newer agents (e.g., Devin) to interact with existing PRs while ensuring complete discussions. For each selected repository, we extracted all PR review threads and associated comments using the GraphQL API.

For PRs containing at least one agent-generated comment, we collected detailed metadata for all inline code comments in the threads following the previous study~\cite{widyasari2025explaining}. 
This metadata includes the author identity, the associated coding agent, the comment content, the resolution status, and any subsequent replies from human developers. In total, this process yielded $54{,}791$ code review comments. However, due to very low representation of Claude and Devin (28 and 50, respectively), we exclude these agents from comparative analysis, as the sample sizes are insufficient for reliable statistical inference. Finally, our dataset includes $54{,}713$ agent-generated comments from Copilot, Cursor, and Codex across $341$ repositories. Table~\ref{tab:comment-resolution} presents their distribution across agents.
\begin{table}[t]
    \centering
    \caption{Review comments taxonomy and description.}
    \resizebox{1.05\columnwidth}{!}{
        \begin{tabular}{p{0.25\columnwidth}p{0.90\columnwidth}}
            \toprule
            \textbf{Category } & \textbf{Description} \\ 
            \midrule
            \functionalmark~Functional Defect & A functionality is missing or implemented incorrectly, which often requires additional code or major fixes. \\ 
            \functionalmark~Validation & Issues with invalid value detection and data sanitization. \\ 
            \functionalmark~Logical & Issues involving comparison, control flow, computation, or other logical errors. \\ 
            \functionalmark~Interface & Issues arising from external components such as libraries, hardware devices, database, or operating system. \\ 
            \functionalmark~Resource & Issues related to the initialization, handling, or release of variables, memory, files, and databases. \\
            \functionalmark~Support & Issues related to support systems, libraries, or their configurations. \\ 
            \functionalmark~Timing & Issues caused by improper thread synchronization when accessing shared resources. \\ 
            \midrule
            \evolvemark~Solution Approach & Suggestions for alternate implementations (e.g., algorithms, data structures). \\ 
            \evolvemark~Documentation & Suggestions to improve code comments or documentation. \\ 
            \evolvemark~Organization of Code & Suggestions for structural refactoring, such as collapse hierarchy, extract superclass, and inline function. \\ 
            \evolvemark~Alternate Output & Suggestions for enhancing error messages, alerts, toast notifications, or function return values. \\ 
           \evolvemark~Naming Convention & Suggestions for renaming software elements to comply with conventions. \\
            \evolvemark~Visual Representation & Suggestions for enhancing code readability, such as adjusting indentation, removing unnecessary whitespace, or reorganizing code. \\ 
            \midrule 
            \discussmark~Question & Questions to understand design and implementation choices. \\ 
            \discussmark~Design Discussion & High-level discussion about design choices, design patterns, and software architecture. \\ 
            \bottomrule[1pt]
            \multicolumn{2}{p{\columnwidth}}{\footnotesize Note: We adopt the code review comment categories and their descriptions from prior works~\cite{ mantyla2008types, turzo2024makes}. \functionalmark~Functional, \evolvemark~Evolvability, \discussmark~Discussion}
        \end{tabular}
    }
    \label{tab:review-taxonomy}
    \vspace{-0.5cm}
\end{table}
%

\textbf{RQ1:} \label{RQ1:methodology}
To answer RQ\textsubscript{1}, we compute the proportion of resolved agent-generated comments across different coding agents and analyze how resolution likelihood varies across comment categories. We define a code review comment as \textit{resolved} when a project collaborator (the PR author, reviewer, or maintainer) explicitly marks the review thread as resolved on GitHub, as indicated by the \texttt{isResolved} field in the GitHub API response~\cite{github_graphql}, along with the identity of the resolver captured in the \texttt{resolvedBy} field.

\textit{Review Comment Classification:}
%
To classify the agent-generated code review comments, we adopted an established taxonomy from prior works~\cite{mantyla2008types, turzo2024makes, lin2025leveraging}. 

We conducted experiments with three LLMs (i.e., Qwen-3-8B, Llama-3.1-70B and GPT-4o) to select our automated annotator. LLMs have shown effectiveness in large-scale annotation tasks and these models have been used widely in code review comment categorization tasks in the literature \cite{goldman2025types, sghaier2025harnessing, nguyen2025exploring, takerngsaksiri2025human}. 
We used JSON-based prompts to ensure consistent annotation. The prompt was chosen through multiple refinement iterations following a trial-and-error approach based on observations. The initial prompt simply instructed the LLM to select the most appropriate category based on the definitions without providing an additional concept. However, it was generating unstructured and verbose outputs in inconsistent formats. To mitigate this issue and to improve performance, we followed \cite{goldman2025types} and employed a justification prompt, encouraging the model to explain its classification. This implicitly triggers Chain-of-Thought (CoT) reasoning, which has been shown to improve the LLM performance in classification tasks \cite{deng2023implicit}. Including justification allowed us to evaluate whether the model is reasoning about relevant features or guessing thereby informing prompt refinement. We also extended the prompt to include a confidence score (ranging from 0 to 1) for each prediction. Full prompt is provided in the replication package \cite{replication_package}.

To assess the soundness of the LLM-based judge, the first two authors conducted a sanity check by manually annotating a random sample of 100 code review comments following previous studies \cite{goldman2025types, sghaier2025harnessing}. Each author independently assigned labels according to the schema defined in Table~\ref{tab:review-taxonomy}. We measured inter-annotator agreement using \textit{Cohen’s kappa} \cite{mchugh2012interrater}, which indicated near-perfect agreement $(\kappa = 0.86)$. Disagreements were subsequently resolved through discussion until consensus was reached. This process provided a reliable human baseline for evaluating the LLM-based annotations.
%
We then compared the LLM-generated labels with the finalized human annotations for each of the models. Our observed $\kappa$ scores for Qwen-3-8B, Llama-3.1-70B and GPT-4o was 0.38, 0.74 and 0.70, respectively. Although Llama-3.1-70B and GPT-4o achieved comparable agreement, we selected Llama-3.1-70B due to its strong performance and practical advantages over proprietary models. The resulting $\kappa$ of 0.74 indicates substantial agreement with human annotations, consistent with prior studies on LLM-based annotation~\cite{zheng2023judging, sghaier2025harnessing}. We therefore used Llama-3.1-70B to annotate the remaining comments.


\textbf{RQ2:} \label{sec:RQ2:methodology}
To answer RQ\textsubscript{2}, we analyze how developer experience affects the resolution of agent-generated comments 
and identify discussion patterns in unresolved comments with open card sorting.

\textit{Developer Experience:}
To ensure consistency, we restrict our resolution analysis to cases where the resolver is also the PR author, 
who resolved at least one agent-generated comment. 
For each repository independently, we categorize developers into \textit{core} and \textit{peripheral} contributors based on their historical activity, following prior work~\cite{mockus2002two, joblin2017classifying, terceiro2010empirical}, where the top 20\% by experience are considered core, while the rest are peripheral. Applying the threshold within each repository rather than globally  normalizes developer experience relative to each project's own activity level, ensuring classifications are not distorted by differences in project sizes or overall PR volume.
Prior research has measured developer experience using various contribution metrics, including the number of commits \cite{eyolfson2011time, posnett2013dual}, lines of code contributed \cite{mockus2002two}, and developer seniority (i.e., years since the first contribution) \cite{mockus2010organizational}. However, commit-based measures have limitations, as they can reflect individual development styles~\cite{robbes2013using}. Thus, prior study~\cite{robbes2013using} recommends using broader contribution activities.
Following this guideline, and recognizing that pull-based development has become the dominant paradigm in open-source software (OSS) projects \cite{gousios2014exploratory}, we adopt a PR-based metric, measuring experience as the number of authored and reviewed closed PRs following prior studies \cite{kononenko2016code, cynthia2026we}. This results in $1,538$ core and $2,554$ peripheral developers in our dataset. Across repositories, the median number of core developers is 3 (mean: 5.4) and the median number of peripheral developers is 2.5 (mean: 9.5).

\textit{Discussion Analysis:}
%
Next, we aim to analyze the discussion patterns that emerged when agent-generated code review comments remained unresolved. Of the $15,652$ unresolved comments in our dataset, $1,056$ received replies, including $655$ from PR authors. We identified $431$ comments where core developers replied and $224$ comments where peripheral developers replied to the comments made by agents.
We then selected a representative sample of cases from each developer group using stratified sampling (95\% confidence level, 5\% margin of error) and also ensured the representation across different agents. Our analysis included 284 comments with core developer replies (Copilot: 168/295, Cursor: 80/100, and Codex: 36/36) and 186 comments with peripheral developer replies (Copilot: 89/114, Cursor: 57/66, and Codex: 40/44). 

After that, we coded the review discussions to identify key concerns raised by developers regarding agent-generated comments. Following prior studies \cite{guzzi2013communication, bacchelli2013expectations, hirao2020code}, we applied open card sorting \cite{spencer2009card} to construct a taxonomy of discussion themes through a bottom-up process \cite{xu2020reinventing}. Two authors independently performed initial coding of the data, resolved disagreements through discussion, and grouped labels into higher-level categories. Inter-rater reliability on a sample of 200 comments using Cohen’s Kappa showed near-perfect agreement ($\kappa = 0.85$). 
The rest of the comments were then independently coded by both authors and resolved through discussion to reach a single agreed-upon label per comment.
%

\textbf{RQ3:} \label{sec:RQ3:methodology}
To answer this RQ, we analyze several characteristics of agent-generated code review comments. The variables considered include the presence of code suggestions, comment length, relevance, clarity, conciseness, and the explanation category. The definition and extraction process for each variable are described in the following subsections. We then analyze how these characteristics relate to the perceived usefulness of a comment. Prior studies commonly use whether a review comment leads to code changes as a proxy for its usefulness and effectiveness~\cite{bosu2015characteristics, goldman2025types}. Following this, we define \textit{usefulness} based on human response: a comment is considered useful if it is resolved by a human reviewer. To ensure usefulness reflects human judgment, we exclude resolutions performed solely by AI agents. %
In addition, building on our RQ2 analysis (Section~\ref{sec:RQ2:methodology}), we also treat comments as useful when developers explicitly state that the issue has been addressed or resolved through subsequent code changes (e.g., ``fixed in commit \textit{$<$commit sha$>$}'', ''done'', ``updated''), even if the comment is not formally marked as resolved. 
After excluding comments that are resolved by the AI agent, we obtain 53,086 labeled comments, including 37,512 useful and 15,574 non-accepted (i.e., not acted upon by human developers).

We perform analyses to compare the distribution of each variable between useful and non-accepted comments. We compute proportions and apply the \textit{Mann-Whitney U test} \cite{fay2010wilcoxon} for continuous variables and the \textit{Chi-Squared test} \cite{mchugh2013chi} for categorical variables to check whether the differences are statistically significant. Furthermore, to understand the combined effect of multiple characteristics, we build a logistic regression model with resolution status as the dependent variable and the extracted code review comment characteristics as the independent variables. 
To account for differences in problem types, we distinguish between functional and evolvability groups~\cite{mantyla2008types, turzo2024makes} (Table~\ref{tab:review-taxonomy}), as different issues may require different types of feedback.
These methods allow us to identify which factors are more related to ``useful'' code review comments.

\textit{Code Suggestion:}\label{sec:code-suggestion}
To identify the presence of code suggestions, we check whether a comment contains the \texttt{suggestion} syntax (i.e., \texttt{``suggestion <code> ''}), which indicates a suggested code change that can be directly applied on GitHub~\cite{github_suggested_changes}. We include this variable because prior studies have shown that comments with suggested code changes are more likely to be adopted and can lead to faster resolution during code review~\cite{palvannan2023suggestion, frommgen2024resolving}.

\textit{Comment Length:}
We measure comment length as the number of characters in each review comment.
Prior studies suggest that useful comments often contain actionable feedback and rich information~\cite{bosu2015characteristics, rahman2017predicting}. Thus, comment length can serve as a proxy for detail and comprehensiveness, although excessively long comments may reduce readability, indicating a trade-off with conciseness. 

\textit{Relevance, Clarity, \& Conciseness:}
Following prior studies~\cite{sghaier2025harnessing, rani2023decade}, each dimension is measured on a scale from 1 to 10, providing a structured way to evaluate the effectiveness of comments. \textit{Clarity} captures how effectively a comment communicates its message, with higher scores indicating clearer communication. \textit{Relevance} measures how well the comment pertains to the code change under review, while \textit{conciseness} assesses whether the comment is brief and focused without unnecessary detail. These scores are assigned using an LLM-as-a-judge approach with LLaMA-3.1-70B during annotation, following the same validation procedure used for comment categorization (Section~\ref{RQ1:methodology}).
\begin{table}[]
\caption{Explanation category and description}\label{tab:explanation-cat}
\resizebox{0.98\columnwidth}{!}{
\begin{tabular}{p{0.11\textwidth}p{0.39\textwidth}}
\toprule[1pt]
\textbf{Exp. Category} & \textbf{Description    }                                                                                                                                                                         \\ \midrule
    Rule                 & Explains by citing a rule, principle, convention, guideline, or standard that should be   followed.                                      \\ \midrule
    Example              & Explains by pointing to a similar implementation, existing code pattern, or reference example. 
    \\ \midrule
    
    Scenario             & Explains by describing a scenario, condition,   edge case, or `what if' situation. 
    \\ \midrule
    Future Implication   & Explains by mentioning what may happen in the future, such as future features, future maintenance, or later changes that would affect this code.          \\ \midrule
    Subjective Opinion   & Explains the feedback as a personal opinion, taste, or preference about the code under review. 
    \\ \midrule
    Issue                & Explains by stating a current issue, fact, root cause, or problematic aspect of the code under review.                                \\ \midrule
    Benefit              & Explains by stating the benefit of the suggested change, such as clearer, simpler, 
    safer, or more maintainable.   \\
    \bottomrule
\end{tabular}}
\end{table}
%

\textit{Explanation Categories:}
Previous studies have highlighted that including explanations in code review comments is important for helping authors understand the feedback~\cite{google_guide, rahman2022example, widyasari2025explaining}. Despite this, we still lack evidence on how AI agents explain their feedback in practice. To address this, we analyze the distribution of explanation types using the taxonomy proposed by Widyasari et al.~\cite{widyasari2025explaining}, which defines seven categories (Table~\ref{tab:explanation-cat}). We use an LLM to (i) determine whether each comment includes an explanation and (ii) assign the corresponding explanation type(s). We do not reuse the prior dataset as ground truth because it contains substantially shorter comments and reports only a single ``main'' explanation type per comment. In contrast, our dataset contains longer comments on average (190 vs.\ 790 characters), making it more plausible that a single comment contains multiple explanation categories.
To validate the LLM-based annotation, we randomly sampled 100 comments and had two authors independently label them~\cite{goldman2025types}. Unlike the prior work~\cite{widyasari2025explaining}, we allowed annotators to assign all applicable explanation types per comment. We use the Jaccard index~\cite{ivchenko1998jaccard} as it allows multi-label measurements of inter-rater agreement and shows a high level of agreement ($J=0.88$). 
 
For automated labeling, we used \textit{LLaMA-3.1-70B}. For each comment, we first prompted the model to identify whether any explanation was present. If so, the model was instructed to assign all explanation categories supported by the comment and to justify each assigned label with: (1) evidence text from the comment, (2) a brief reasoning, and (3) a confidence score $(0-1)$. By requiring evidence and reasoning, it facilitates error analysis and aligns with prior findings that rationales can improve performance~\cite{deng2023implicit}. To increase precision, we retained only labels with confidence $\ge 0.9$. We compared the resulting LLM labels against the human annotation and observed high agreement ($J=0.90$), indicating near-perfect consistency. Given this performance, we used the same procedure to annotate the remaining comments in the dataset.
%
\section{Resolution Patterns Across Agents and Comment Types (RQ1)} \label{sec:RQ1-result}

In this section, we present the resolution frequency for each coding agent and how resolution likelihood varies by comment types. Table~\ref{tab:comment-resolution} shows clear differences in both volume and adoption across agents. Copilot dominates and achieves the highest resolution rate (72.9\%), indicating strong developer acceptance. Cursor shows a slightly lower but comparable rate, while Codex has the lowest resolution rate, suggesting lower adoption. Overall, Copilot dominates both in developer adoption and the number of resolved comments.

%
\begin{table}[t]
  \centering
  \caption{Agent-generated comment resolution frequency for each coding agent}
  \label{tab:comment-resolution}
  \resizebox{\columnwidth}{!}{
      \begin{tabular}{
          >{\raggedright\arraybackslash}p{1.9cm} 
          >{\raggedleft\arraybackslash}p{1.2cm}       
          >{\raggedleft\arraybackslash}p{2.8cm}     
          >{\raggedleft\arraybackslash}p{1.3cm}                          
        }
        \toprule
        \textbf{Coding agent} & \textbf{\#Comments}   &   \textbf{\#Resolved Com.} &    \textbf{Res. rate}\\
        \midrule
        Copilot & 45,668    &   33,265  &   72.9\% \\
        Cursor  & 6,778     &   4,554   &   67.2\%\\
        Codex   & 2,267     &   1,242   &   54.8\%\\
        \bottomrule
      \end{tabular}
    }
\end{table}
\begin{figure}
    \centering
    \includegraphics[width=1\linewidth]{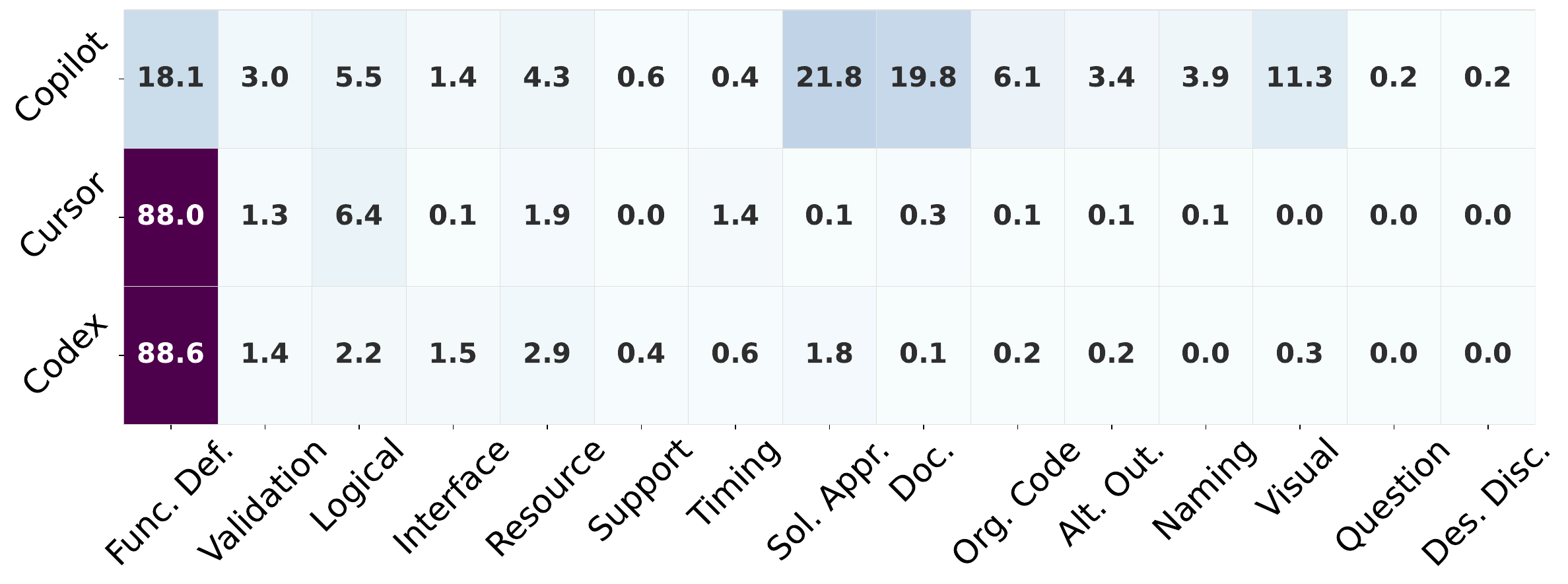}
    \caption{Distribution (\%) of resolved agent-generated comments by comment type across agents} 
    \label{fig:comment-type}
\end{figure}
Next, Fig.~\ref{fig:comment-type} presents the distribution of resolved agent-generated comments across different comment categories and coding agents. Among the agents, Copilot exhibits a relatively balanced distribution of resolved comments across multiple categories. For instance, Copilot most frequently comments on \textit{solution approaches} (7238/33265, 21.8\%), followed by documentation (6580/33265, 19.8\%), and functional defect issues (6030/33265, 18.1\%). Visual representation and organisation of code also contribute notably. This suggests that Copilot acts as a more general-purpose reviewer, addressing not only functional issues but also code structure, clarity, and design.
In contrast, Codex and Cursor show a strong concentration of resolved comments related to \textit{functional defects}. These comments achieve the highest resolution rates across agents, particularly for Codex (1100/1242, 88.6\%) and Cursor (4006/4554, 88\%). Compared with Copilot, the distribution of other comment categories for these agents is highly uneven. For example, the second most frequent category for Codex is \textit{resource} issues, accounting for only 2.9\% (36/1242) of its resolved comments, while for Cursor the second most frequent category is \textit{logical} issues at 6.4\% (293/4554).
Overall, the distribution highlights notable differences in agent review behavior, suggesting that some agents emphasize broader review coverage while others specialize in detecting functional defects. 

\begin{center} 
{\setlength{\fboxsep}{6pt}
\colorbox{blue!5!white}{%
  \parbox{0.95\linewidth}{%
    \textbf{Answer to RQ1: } Copilot contributes the most comments and over 72\% of the comments are resolved, with a relatively balanced distribution across categories, while Codex and Cursor focus primarily on functional defects (88\%). 
  }%
}}
\end{center}
%
%
\section{Developer Experience and Unresolved Discussion Patterns (RQ2)} \label{sec:RQ2-result}
\begin{figure}
    \centering
    \includegraphics[width=\linewidth]{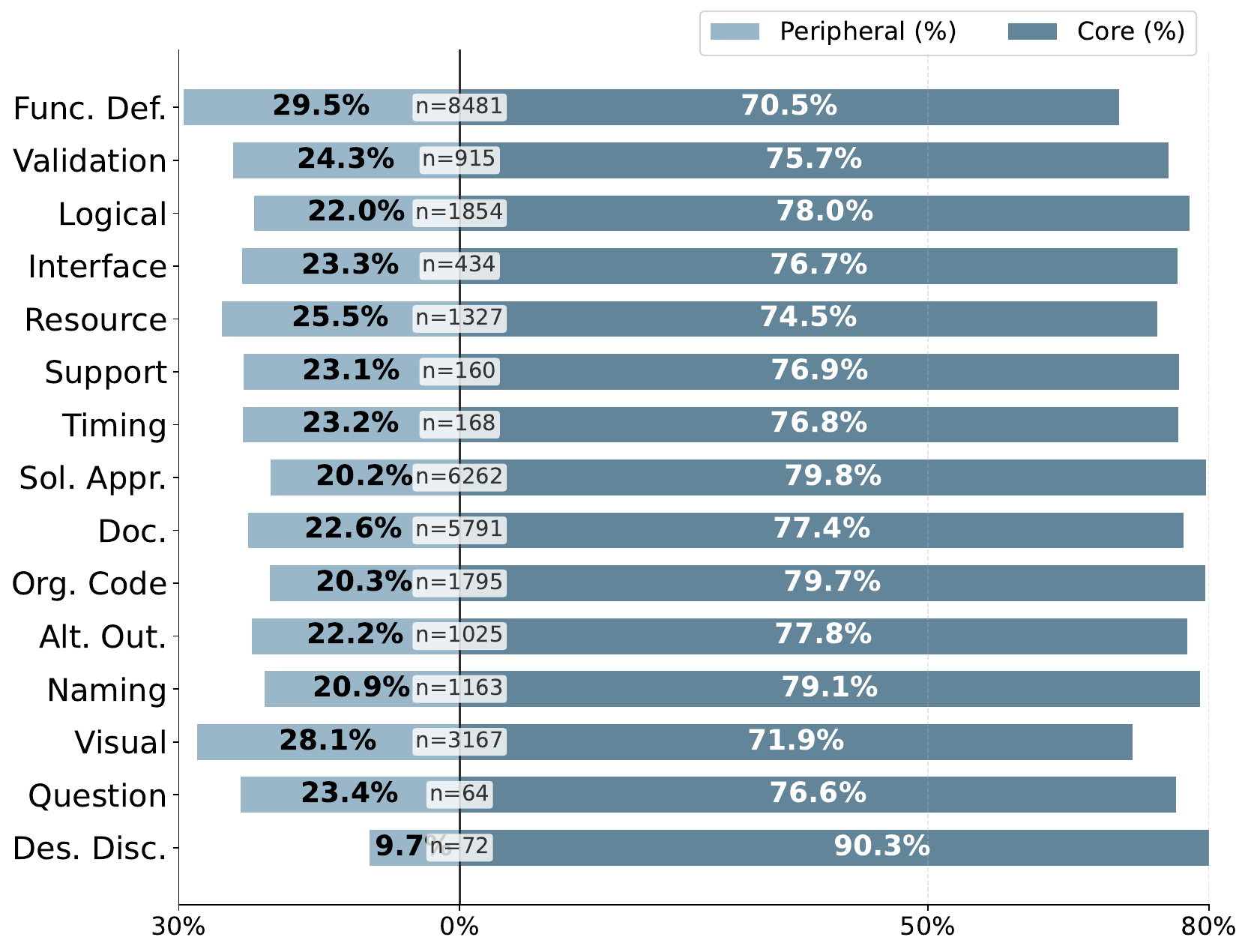}
    \caption{Distribution of resolved comments per comment types for core and peripheral developers.}
    \label{fig:comment-type-core-peripheral}
\end{figure}

In this section, we investigate how developer experience influences the acceptance of comments and the discussion patterns that emerge when they remain unresolved. 
\begin{table}[t]
  \centering
  \caption{Distribution of resolved comments by PR author type (core vs. peripheral) across agents.}
  \label{tab:agent-core-peripheral}
  \resizebox{0.98\columnwidth}{!}{
          \begin{tabular}{p{0.15\columnwidth}p{0.22\columnwidth}p{0.22\columnwidth}p{0.24\columnwidth}} 
        \toprule
        \textbf{Coding agent} & \textbf{\#Resolved Comments}   &   \textbf{Core devs.} &    \textbf{Peripheral devs.}\\
        \midrule
        Copilot & 29,108    &   78.1\%   &   21.9\% \\
        Cursor  & 2,714     &   54.6\%   &   45.4\%\\
        Codex   & 866       &   58.4\%   &   41.6\%\\
        \bottomrule
      \end{tabular}
    }
\end{table}
Table~\ref{tab:agent-core-peripheral} presents the distribution of resolved agent-generated review comments across developer groups for different coding agents. Although core and peripheral groups are similar in size, core developers resolve a substantially larger share of agent-generated comments across most agents. For instance, 78.1\% of Copilot-generated comments are resolved by core developers, suggesting that experienced developers are primarily responsible for incorporating automated review suggestions into the codebase. In contrast, Cursor and Codex show a more balanced distribution between core and peripheral developers, indicating that feedback from these agents is addressed by developers with varying levels of experience.
To further examine the role of developer experience, we analyze how resolved agent-generated comments vary across different comment categories. As shown in Fig.~\ref{fig:comment-type-core-peripheral}, core developers consistently resolve the majority of comments across all categories, particularly for suggestions related to the solution approach, documentation, naming convention, organization of code, and design discussion, which often require deeper familiarity with project design and coding standards. However, functional defect comments have a relatively higher share of peripheral developer involvement compared to other categories.
These findings indicate that core developers tend to take primary responsibility for integrating agent-generated comments related to evolvability, while peripheral developers contribute more frequently when resolving defect-oriented suggestions.
%
\begin{center} 
{\setlength{\fboxsep}{6pt}
\colorbox{blue!5!white}{%
  \parbox{0.95\linewidth}{%
    Core developers resolve the majority of comments, especially design and structural changes, while peripheral developers are relatively more involved in functional defect fixes.
  }%
}}
\end{center}

To further understand how developers respond to agent-generated feedback when comments remain unresolved, we analyze the discussion patterns that emerge during follow-up interactions. Table~\ref{tab:discussion-taxonomy} presents the resulting taxonomy with 10 high-level categories and 13 sub-categories, which we define in the following. 

\begin{table}[t]
\centering
\caption{Discussion taxonomy and distribution across developer groups.}
\resizebox{\columnwidth}{!}{
\begin{tabular}{p{4.5cm}p{0.8cm}p{0.8cm}p{1cm}}
\toprule[1pt]
\textbf{Category-Subcategory} & \textbf{\#Disc.$^{a}$} & \textbf{Core} & \textbf{Periph.} \\
\midrule
\textbf{Intentional Design Decision} & 112 {\scriptsize(55/48/9)} & 81 & 31 \\
\quad Context-Specific Implementation & 64 & 44 & 20 \\
\quad Developers' Preference & 11 & 11 & 0 \\
\quad Expected Behaviour & 37 & 26 & 11 \\

\midrule


\textbf{Incorrect Suggestion} & 67 {\scriptsize(36/19/12)} & 32 & 35 \\
\quad Factually Wrong or False Positive & 63 & 29 & 34 \\
\quad Agent Hallucinated & 4 & 3 & 1 \\

\midrule


\textbf{Challenging Agent Feedback} & 39 {\scriptsize(22/12/5)} & 25 & 14 \\
\quad Disagreement with Suggestions & 27 & 14 & 13 \\
\quad Questioning Agent's Claim & 12 & 11 & 1 \\

\midrule

\textbf{Suggestions Deferred} & 25 {\scriptsize(12/9/4)} & 16 & 9 \\
\quad Future Improvement & 21 & 15 & 6 \\
\quad Deferred to Future PR & 4 & 1 & 3 \\

\midrule

\textbf{Delegation of Work} & 31 {\scriptsize(22/6/3)} & 24 & 7 \\
\quad Agent Task Delegation & 23 & 18 & 5 \\
\quad Delegating to Developer & 8 & 6 & 2 \\

\midrule

\textbf{Dismissed as Low-Value} & 11 {\scriptsize(7/4/0)} & 8 & 3 \\ \hline
\textbf{Needs Further Discussion} & 36 {\scriptsize(19/10/7)} & 20 & 16 \\ \hline
\textbf{Acceptable Trade-Offs} & 19 {\scriptsize(9/10/0)} & 13 & 6 \\ \hline
\textbf{Missed Existing Fix} & 12 {\scriptsize(9/2/1)} & 9 & 3 \\ 

\midrule


\textbf{Accepted Agent Feedback} & 114 {\scriptsize(66/31/17)} & 54 & 60 \\
\quad Accepted Suggestion & 78 & 35 & 43 \\
\quad Addressed Afterwards & 36 & 19 & 17 \\

\bottomrule[1pt]
\multicolumn{4}{l}{{\scriptsize $^{a}$Agent breakdown in parentheses: Copilot / Cursor / Codex}}
\end{tabular}
}
\vspace{-0.5cm}
\label{tab:discussion-taxonomy}
\end{table}
\textbf{\textit{1. Intentional Design Decision.}}
This category includes instances where developers deliberately reject agent-generated suggestions because the existing implementation aligns with system design, contextual requirements, or development practices. 

\textit{A. Context Specific Implementation:} Developers reject suggestions when code is tailored to project-specific constraints not visible to the agent, rendering the proposed changes inappropriate given the system’s design assumptions. For example, in \textcolor{Mycolor}{\textit{PR\#3205}}~\cite{opik_pr_3205}, the agent suggested a different logging approach, but the developer clarified it to be intentional, stating that \textit{``No, the idea is to use add logs to the docker stream.''} 

\textit{B. Developers' Preference:} This sub-category includes instances where developers reject suggestions based on personal or team-level coding preferences. For example, in \textcolor{Mycolor}{\textit{PR\#2854}}~\cite{azure_pr_2854}, the developer preferred runtime errors over null handling, by stating, \textit{``We discussed and think that null-forgiving is better in this case since we'd like the runtime error instead.''} 

\textit{C. Expected Behaviour:} Developers reject suggestions when the current behavior is already correct and intentional. For example, in \textcolor{Mycolor}{\textit{PR\#105249}}~\cite{sentry_pr_105249}, the agent flagged delayed alert disappearance, but the developer clarified that this is intentional by stating, \textit{``Desired behaviour, only shouldn't exist on next load.''} 

\textbf{\textit{2. Incorrect Suggestion.}}
This category captures instances where developers reject agent-generated review comments because the suggestions are incorrect or based on flawed reasoning.

\textit{A. Factually Wrong or False Positive:} 
The agent either incorrectly flags a non-existent issue (false positive) or provides a factually incorrect suggestion due to misunderstanding the code, system behavior, or APIs. In such cases, developers clarify that no change is required or reject the suggestion by correcting the inaccurate claim. For example, in \textcolor{Mycolor}{\textit{PR\#42779}}~\cite{posthog2024pr42779}, the agent raised a grammar issue, but the developer clarified by stating, \textit{``False positive - the initial message already says "1 commit, 1 workflow" (singular).''} 


\textit{B. Agent Hallucination:} 
Agents make suggestions based on nonexistent or fabricated assumptions about the code. These suggestions often introduce irrelevant or non-existent changes that do not align with the actual implementation, leading developers to reject them. In \textcolor{Mycolor}{\textit{PR\#54198}}~\cite{ray_pr_54198}, the agent claimed that required Java setup steps were removed and would break the build, but the developer dismissed this as incorrect, indicating the steps never existed.

\textbf{\textit{3. Challenging Agent Feedback.}}
This category includes instances where developers actively question the necessity, relevance, or reasoning behind agent-generated feedback, often expressing skepticism or requesting clarification. 

\textit{A. Disagreement With Suggestion:} 
Developers either explicitly disagree with the agent’s recommendation or acknowledge it but consider it unnecessary to implement. For example, in \textcolor{Mycolor}{\textit{PR\#56790}}~\cite{ray_pr_56790}, the agent flagged a timing issue, but the developer disagreed with the suggestion, stating, \textit{``I don't think there's a bug here''}. 

\textit{B. Questioning Agent's Claim:} This sub-category refers to cases where developers question the validity or reasoning behind the agent’s feedback, often expressing uncertainty or seeking further clarification. For instance, in \textcolor{Mycolor}{\textit{PR\#56752}}~\cite{ray_pr_56752}, the agent warned about caching issues, and the developer responded by questioning the claim and asked the agent to re-verify stating, \textit{``are you sure? check where this docker file is used in the original code base, and tell me how the caching could be enabled.''} 


\textbf{\textit{4. Suggestions Deferred.}}
This category captures instances where developers acknowledge agent-generated suggestions as valid or potentially useful but postpone the implementation. 

\textit{A. Future Improvement:} 
Developers recognize the value of the suggestion but defer it as a potential enhancement to be addressed in the future, without committing to a specific timeline. For example, in \textcolor{Mycolor}{\textit{PR\#104594}}~\cite{sentry_pr_104594}, the agent highlighted empty bulk action buttons, developer acknowledged this and stated it would be implemented later once more information is available, stating, \textit{``yup. will revisit this when i have info about the bulk endpoints.''}. 

\textit{B. Deferred to Future PR:} 
Developers explicitly state that the suggestion will be handled in a separate or upcoming pull request, indicating a structured plan to implement the change at a later stage. For example, in \textcolor{Mycolor}{\textit{PR\#6568}}~\cite{onyx_pr_6568}, the agent flagged an API issue, and the developer stated that it will be fixed in a future pull request, stating, \textit{``This will be addressed in a follow-up PR that clarifies 401 403 semantics in our API''}.

\textbf{\textit{5. Delegation of Work.}} 
This category captures instances where developers do not directly act on agent-generated suggestions but instead delegate the task to others. 

\textit{A. Agent Task Delegation:} 
Developers assign the task back to an automated agent or tool, expecting it to handle or reprocess the suggested change. For example, in \textcolor{Mycolor}{\textit{PR\#1812}}~\cite{roboflow_pr_1812}
 and \textcolor{Mycolor}{\textit{PR\#17152}}~\cite{litellm_pr_17152}, developers asked the agent to apply or fix the issue. 

\textit{B. Delegating to Another Developer:} 
Developers assign the responsibility of addressing the suggestion to another team member, indicating task handoff within the team. In \textcolor{Mycolor}{\textit{PR\#2333}}~\cite{kombu_pr_2333}, the developer asked another member for input before acting on the suggestion.

\textbf{\textit{6. Dismissed as Low-Value:}}
This category captures cases where developers dismiss agent suggestions as low-quality or irrelevant, often with negative or sarcastic responses, indicating that they do not find the suggestion useful. For example, in \textcolor{Mycolor}{\textit{PR\#5324}}~\cite{fastdeploy_pr_5324}, the developer replied to the agent comment by stating, \textit{``Comment is not helpful''} while in \textcolor{Mycolor}{\textit{PR\#1448}}~\cite{smolagents_pr_1448}, the developer rejected the agent's suggestion by replying sarcastically, stating, \textit{``No it isn't, go home copilot you're drunk''}. 

\textbf{\textit{7. Needs Further Discussion.}}
This category captures cases where agent-generated feedback initiates discussion among developers, requiring additional analysis or input and leaving the issue unresolved in the current thread. In \textcolor{Mycolor}{\textit{PR\#2173}}~\cite{openai_agents_pr_2173},  the agent noted that recent changes may remove important records, the developer asked for suggestions on how to address it, stating, \textit{``You're right, we need to handle this. Any suggestions?''} 

\textbf{\textit{8. Acceptable Trade-offs.}}
This category captures cases where developers acknowledge the agent’s suggestion but choose not to apply it because the current implementation reflects an intentional trade-off, such as balancing performance, readability, or complexity. For example, in \textcolor{Mycolor}{\textit{PR\#103695}}~\cite{sentry_pr_103695}, the agent pointed out a display issue, but the developer responded that this is acceptable for now, stating, \textit{``this is fine for now since we haven't finalized what it's going to look like in the success page (see TODO)''}. 

\textbf{\textit{9. Missed Existing Fix}}
This category captures cases where the agent suggests a change or identifies an issue that has already been addressed in the code. Developers respond by pointing out that the fix was already handled, indicating that the agent overlooked the current implementation. For example, in \textcolor{Mycolor}{\textit{PR\#19442}}~\cite{mlflow_pr_19442}, the agent warned about a potential input issue, but the developer clarified that it is already handled in the code, stating, \textit{``4 lines above is the check and handling for this use case. It's already handled.''} 

\textbf{\textit{10. Accepted Agent Feedback.}}
This category captures cases where developers accept and act on agent suggestions but do not mark the comment as resolved. 

\textit{A. Accepted Suggestion:}  
Developers explicitly agree with the agent's suggestion and implement the recommended changes. Responses typically indicate clear acceptance through brief confirmations or direct fixes, such as \textit{``Fixed'', ``Made the change'' and ``Done''}. 


\textit{B. Addressed After Agent's Comment:} 
Developers act on the agent’s feedback by applying the suggested changes and either request re-review or indicate that the issue has been resolved in a subsequent commit or PR. For example, in \textcolor{Mycolor}{\textit{PR\#16382}}~\cite{mlflow_pr_16382}, the developer mentioned that the agent's suggestion has been fixed in a follow-up commit. 

\vspace{2pt}
\noindent Additionally, we identified 4 discussions that could not be mapped to any of the defined categories, such as comments like ``same''. Excluding \textit{Accepted Feedback}, the most common reason for non-adoption is \textit{Intentional Design Decision}, predominantly from core developers (72\%, 81/112). This suggests that experienced developers often reject suggestions due to deeper project understanding. This is followed by \textit{Incorrect Suggestion}, which is more balanced between core (48\%) and peripheral developers (52\%), likely because identifying incorrect feedback may not require deep project knowledge. These findings highlight that agents still struggle with project context and may produce factually incorrect suggestions.
Nonetheless, agent-level breakdown reveals patterns consistent with RQ\textsubscript{1} as \textit{Intentional Design Decision} is disproportionately prevalent in Cursor discussions (32\%) compared to Copilot (21.4\%) and Codex (15.5\%), suggesting a more pronounced context-awareness gap despite Cursor's functional focus. \textit{Incorrect Suggestion} is slightly elevated for Codex (20.6\%) relative to Copilot (14\%) and Cursor (12.6\%), consistent with Codex's lowest resolution rate in RQ\textsubscript{1}. \textit{Acceptable Trade-Offs} and \textit{Dismissed as Low-Value} appear only in Copilot and Cursor discussions. The remaining categories show no substantial agent-specific variation and are broadly proportional across agents.
\begin{center} 
{\setlength{\fboxsep}{6pt}
\colorbox{blue!5!white}{%
  \parbox{0.95\linewidth}{%
    \textbf{Answer to RQ2: }  
    Core developers resolve most comments, particularly design and structural changes. When unresolved, they rejected over 72\% of comments due to \textit{Intentional Design Decisions}, while both groups similarly encountered \textit{Incorrect Suggestions}, indicating agents often miss context or produce flawed feedback. At the agent-level, \textit{Intentional Design Decision} is most prevalent in Cursor and \textit{Incorrect Suggestion} is most elevated for Codex.
  }%
}}
\end{center}
\begin{table}[t]
    \centering
    \caption{Usefulness rate by the presence of code suggestions}
    \resizebox{1\columnwidth}{!}{
        \begin{tabular}{
          >{\raggedright\arraybackslash}p{2.2cm} 
          >{\raggedleft\arraybackslash}p{2.5cm}       
          >{\raggedleft\arraybackslash}p{2.5cm}  
           >{\raggedleft\arraybackslash}p{2.5cm} 
        }
            \toprule[1pt]
            Code suggestion & All & Func. Group & Evolv. group \\
            \midrule
            No suggestion & 64.6\% (15117/23406) & 61.3\% (8197/13367) & 67\% (6807/9873)\\
            Has suggestion & 75.5\% (22463/29758) & 74.4\% (7056/9485) & 76.1\% (15333/20147)  \\
            \bottomrule[1pt]
        \end{tabular}
    }
    \label{tab:code-suggestion}
\end{table}
%
\section{Characteristics of Useful Agent-Generated Comments (RQ3)} \label{sec:RQ3-result}
While prior analysis highlights the role of developer expertise in shaping the acceptance behavior, it remains unclear which characteristics of the comments themselves contribute to their usefulness. Understanding these properties is essential for designing more effective agent-generated feedback. Specifically, we consider six variables (Section~\ref{sec:RQ3:methodology}) and compare their distribution between comments perceived as useful and those non-accepted by developers.
Our results show that actionability plays a central role in determining the usefulness of agent-generated comments. As shown in Table~\ref{tab:code-suggestion}, comments with inline code suggestions achieve a substantially higher resolution rate (75.5\%) compared to those without suggestions (64.5\%), and this difference is statistically significant (Chi-Squared test, $p < 0.05$, Cram\'{e}r's $V = 0.12$), indicating a small but consistent effect. This pattern holds across both functional and evolvability issue groups (useful rate 74.4\% and 76.1\%, respectively), indicating that directly applicable suggestions reduce the effort required for developers to adopt suggested changes. 
\begin{figure}
    \centering
    \includegraphics[width=0.9\linewidth]{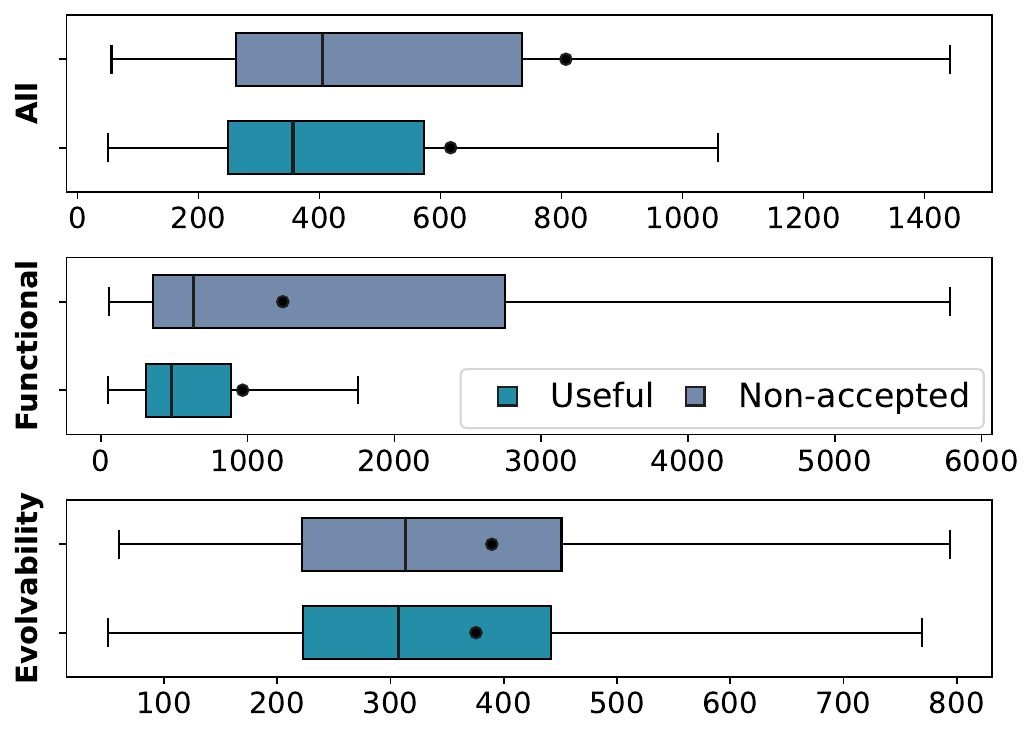}
    \caption{Distribution of comment length for useful and non-accepted comments ($\bullet$ represents the mean value).}
    \label{fig:resolved-unresolved-comment-length}
    \vspace{-0.5cm}
\end{figure}

%
\begin{figure}
    \centering
    \includegraphics[width=\linewidth]{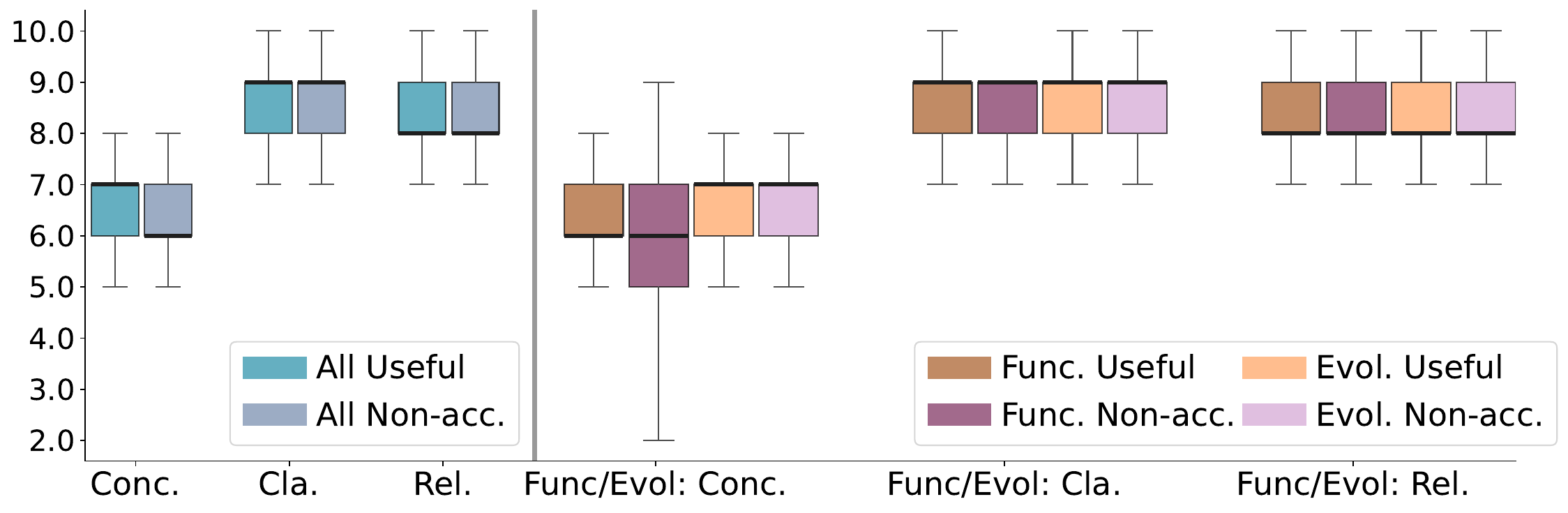}
    \caption{Distribution of conciseness, clarity, and relevance for useful and non-accepted comments.}
    \label{fig:resolved-unresolved-metric-scores}
\end{figure}
Beyond actionability, we examine structural and qualitative properties of comments. The comparison shown in Fig.~\ref{fig:resolved-unresolved-comment-length} of comment length distribution indicates that non-accepted comments tend to be longer on average (mean=807.1, median=405), compared to useful comments (mean=616.6, median=356). A Mann-Whitney U test confirms that this difference is statistically significant ($p < 0.001$). However, the effect size is negligible, indicating that the practical difference between the groups is small. 
This pattern also holds across both functional and evolvability issue groups. The non-accepted comments in the functional group are substantially longer (mean=1239.9, median=634) compared to useful comments (mean=966.3, median=480). The longer comments in the functional group may reflect a tendency for the agents to provide more detailed explanations for functional issues, describing potential runtime behaviors, impacts, and recommended fixes. In contrast, in the evolvability issue group, the difference between non-accepted and useful comments is marginal and non-significant, indicating that comment length has practically no impact on the usefulness of comments in this group.
%

Regarding quality-related metrics, as shown in Fig.~\ref{fig:resolved-unresolved-metric-scores}, conciseness has a slightly higher median for useful comments (7 vs. 6), whereas relevance and clarity share identical medians across both groups. Although we found statistically significant differences, the effect sizes for all three metrics are negligible. 
When analyzed by comment type, both functional and evolvability comments exhibit similar distributions. Functional comments appear slightly more concise than evolvability ones (median: 6 vs.\ 7), reflecting differences in comment nature rather than usefulness. Overall, these findings suggest that writing quality alone does not meaningfully distinguish useful from non-accepted comments.

\begin{figure}
    \centering
    \includegraphics[width=1.1\linewidth]{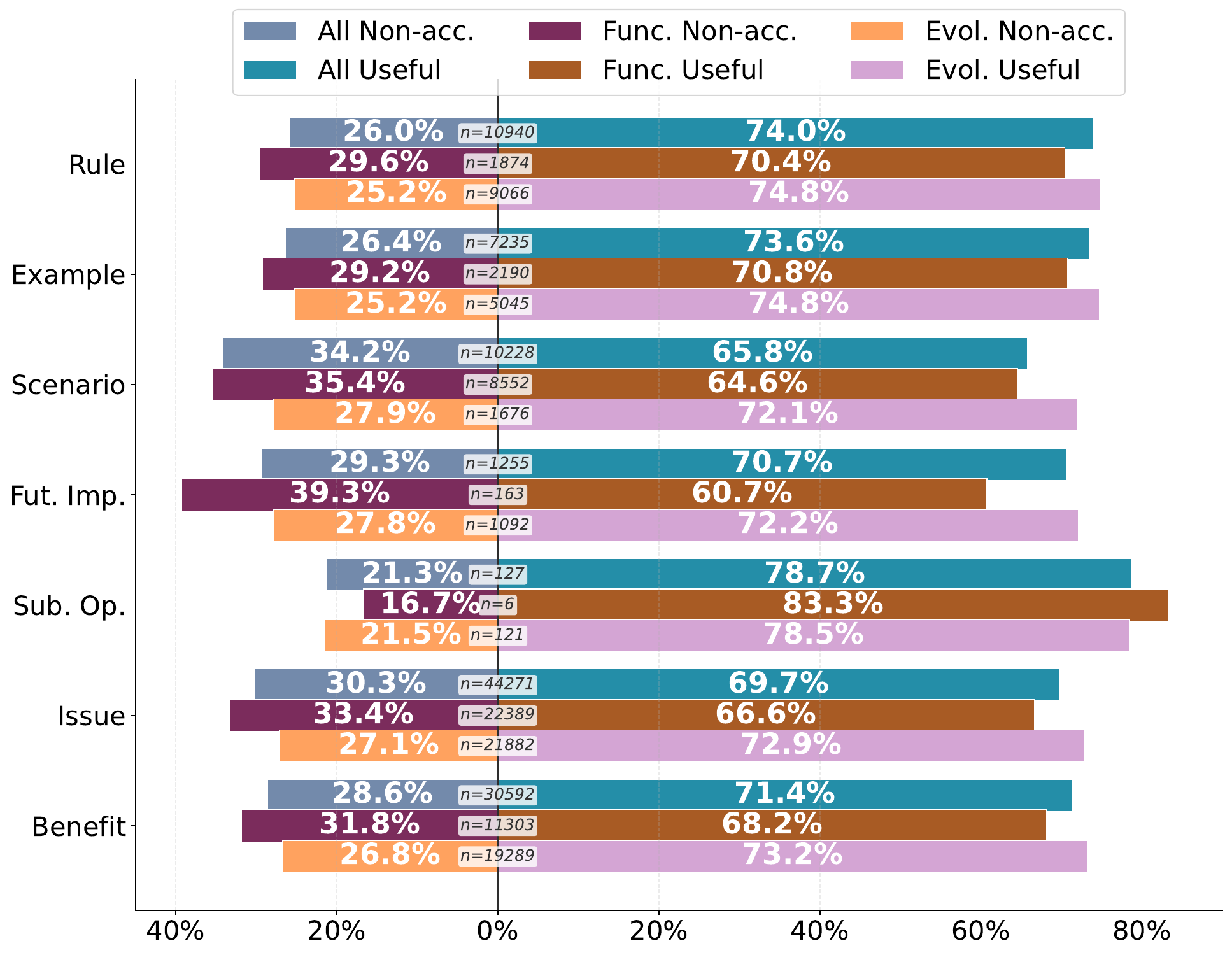}
    \caption{Distribution of useful and non-accepted comments per explanation types.}
    \label{fig:resolved-unresolved-explanation-types}
    \vspace{-0.5cm}
\end{figure}

The analysis of explanation types in Fig.~\ref{fig:resolved-unresolved-explanation-types} reveals that explanations providing concrete guidance, such as rules, examples, benefits, or direct opinions, are associated with higher usefulness rates, whereas scenario and issue-based explanations tend to be less effective. A Chi-Squared test indicates a significant association between explanation type and resolution ($p < 0.001$), with limited practical impact. Scenario-based and rule-based explanations contribute the most to the Chi-Squared statistic, indicating that these categories primarily drive the observed differences. When distinguishing between functional and evolvability groups, a consistent pattern emerges in which evolvability comments achieve higher usefulness rates than functional comments across nearly all explanation types. The gap is most pronounced for future-implication (an increase of 11.5), scenario-based explanations (7.5), and issue-based (6.3), suggesting that evolvability-related feedback benefits from a richer explanatory context. Subjective opinion is the only exception; however, the functional sample size is very small.

To examine explanatory richness, we analyze the number of explanation types per comment (Table~\ref{tab:number-of-explanation}). Comments with two explanation types achieve a higher usefulness rate (71.7\%), followed by single types (68.7\%), with performance declining beyond two types. Interestingly, comments without explanations show the highest resolution rate (83.7\%), but this group is relatively small and largely consists of surface-level categories (89\%) such as documentation (234), visual representation (178), and naming (75). This suggests that the observed effect is driven by category characteristics rather than the absence of explanations.
This is further confirmed by the lower usefulness in the functional group (75\%) than in the evolvability group (84\%), suggesting the high overall rate is driven by straightforward evolvability issues (e.g., documentation, naming). 
For comments with at least one explanation type, the evolvability group consistently outperforms the functional group across all counts. In both groups, the usefulness peaks at two explanation types and declines beyond that, suggesting that while some explanation helps, excessive detail may reduce actionability by introducing unnecessary complexity.
\begin{table}[H]
    \centering
    \caption{Usefulness rate by number of unique explanation types per comment.}
    \resizebox{1\columnwidth}{!}{
        \begin{tabular}{
          >{\raggedright\arraybackslash}p{0.5cm} 
          >{\raggedleft\arraybackslash}p{2.5cm}       
          >{\raggedleft\arraybackslash}p{2.5cm} 
          >{\raggedleft\arraybackslash}p{2.5cm}                         
        }
            \toprule[1pt]
            \textbf{\#Exp.} & \textbf{All} & \textbf{Func. group} & \textbf{Evolv. group} \\
            \midrule
            0 & 83.7\% (547/653)    & 75\% (12/16)   & 84\% (535/637) \\
            1 & 68.7\% (7402/10769) & 63.8\% (3314/5197) & 73.4\% (4088/5572) \\
            2 & 71.7\% (22002/30706) & 68.4\% (8070/11807) & 73.7\% (13932/18899) \\
            3 & 69.3\% (7134/10292) & 66.2\% (3724/5627) & 73.1\% (3410/4665) \\
            $\geq$4 & 65\% (258/397) & 63.3\% (124/196) & 66.7\% (134/201) \\
            \bottomrule[1pt]
        \end{tabular}
    }
    \label{tab:number-of-explanation}
\end{table}

To further understand the combined effect of these characteristics, we build a logistic regression model with usefulness as the dependent variable and the extracted features as independent variables. Results (Table~\ref{tab:logit-coef-pvalue}) show that inline code suggestion is the strongest positive predictor ($OR = 1.62$), indicating that comments with it are more likely to be useful. 
Comment length has a small but significant negative effect ($OR = 0.93$), suggesting that longer comments are marginally less likely to be acted upon, consistent with the univariate finding that useful comments tend to be shorter.
Among explanation types, rule-based explanations, benefits, and examples are positively associated with usefulness, with rule-based opinion showing the largest effect ($OR = 1.14$). In contrast, the presence of explanations has a negative effect, suggesting that concise, action-oriented comments are more readily adopted. Relevance, conciseness, and clarity show small but significant positive effects, indicating a modest role of writing quality.
When analyzed by comment type, the results reveal distinct predictor profiles for functional and evolvability comments. Code suggestion remains the strongest predictor for both groups. For functional comments, a stronger negative effect of comment length is observed ($OR = 0.86$), suggesting the importance of brevity, while relevance and benefit-based explanations show modest positive associations. Interestingly, functional comments do not have a negative effect on the presence of explanation. For evolvability comments, the negative effect of explanation presence is more pronounced ($OR = 0.58$), and notably, rule-based explanations emerge as a significant predictor in this group. Conciseness, relevance, and clarity also show modest positive associations in this group.

Overall, the results indicate that actionable and concise feedback is the primary driver of comment usefulness, with inline code suggestions emerging as the strongest predictor of resolution. However, specific explanation types, contribute positively when they complement actionable suggestions. Furthermore, the differences between functional and evolvability comments suggest that brevity is more critical for functional feedback, whereas writing quality plays a relatively larger role in evolvability-related comments.
Additionally, to examine whether these predictors hold consistently across agents, we conducted separate logistic regression analyses for each agent. The result reveals that the predictors identified in the full model are largely driven by Copilot comments. Since Copilot accounts for 86\% of the dataset, the overall model closely mirrors Copilot's interaction patterns, while also revealing an additional explanation-type effect, which is issue-based explanation. Comments that merely stated an issue remained negatively associated with resolution ($OR = 0.92$). In contrast, the same characteristics exhibited limited predictive power for Cursor and Codex, and agent-specific models fail to converge due to data sparsity and separation issues. These findings suggest that the usefulness of agent-generated review comments is influenced by factors beyond comment characteristics alone. The substantially lower resolution rates for Cursor (67\%) and Codex (53\%) compared to Copilot (73\%), despite controlling for comment characteristics, indicate that agent-specific factors may influence developer responses. Differences in how agents are integrated into the review workflow, as well as agent integration and developer familiarity with these agents, may partly explain why some agent-generated feedback is more likely to be considered useful than others. The results of the agent-specific regression analyses are provided in the replication package~\cite{replication_package}.

\begin{center} 
{\setlength{\fboxsep}{6pt}
\colorbox{blue!5!white}{%
  \parbox{0.95\linewidth}{%
    \textbf{Answer to RQ3: }Including concrete suggestions is the primary driver of usefulness in agent-generated review comments ($OR = 1.609$). Among explanation types, providing a rule, benefit, or example modestly increases usefulness likelihood (OR range: 0.77--0.134), highlighting that the effectiveness of explanations depends more on their type than their mere presence. In contrast, writing quality factors such as clarity and conciseness have minimal impact. The predictive effects are largely driven by Copilot, while agent-specific factors play a greater role for Cursor and Codex.
  }%
}}
\end{center}
\begin{table}[t]
  \caption{Significant logistic regression analysis of comment characteristics and their impact on useful likelihood ($p < 0.05$). OR = odds ratio; CI = 95\% confidence interval.}
  \label{tab:logit-coef-pvalue}
  \small
  \resizebox{\columnwidth}{!}{
      \begin{tabular}{lrrrr}
        \toprule
        \textbf{Predictor} & \textbf{Coef.} & \textbf{OR} & \textbf{95\% CI} & \textbf{$p$} \\
        \toprule
        \multicolumn{5}{c}{\cellcolor[HTML]{C0C0C0}All}
        \\
        Code\_suggestion       &  0.481 & 1.617 & [1.555, 1.682] & ${<}0.0001$ \\
        log\_comment\_length   & -0.077 & 0.926 & [0.898, 0.954] & ${<}0.0001$ \\
        has\_explanation       & -0.522 & 0.593 & [0.474, 0.742] & ${<}0.0001$ \\
        exp\_Rule              &  0.134 & 1.144 & [1.081, 1.210] & ${<}0.0001$ \\
        exp\_Benefit           &  0.084 & 1.087 & [1.043, 1.134] & $0.0001$ \\
        exp\_Examples          &  0.077 & 1.080 & [1.017, 1.148] & $0.0126$ \\
        Relevance              &  0.045 & 1.046 & [1.025, 1.067] & ${<}0.0001$ \\
        Conciseness            &  0.057 & 1.059 & [1.029, 1.090] & $0.0001$ \\
        Clarity                &  0.022 & 1.022 & [1.001, 1.044] & $0.0371$ \\
        \toprule
        \multicolumn{5}{c}{\cellcolor[HTML]{C0C0C0}Functionability issue}
        \\
        Code\_suggestion       &  0.514 & 1.672 & [1.572, 1.778] & ${<}0.0001$ \\
        log\_comment\_length   & -0.156 & 0.855 & [0.817, 0.896] & ${<}0.0001$ \\
        Relevance              &  0.043 & 1.044 & [1.013, 1.075] & $0.0051$ \\
        exp\_Benefit           &  0.064 & 1.066 & [1.005, 1.131] & $0.0337$ \\
        \toprule
        \multicolumn{5}{c}{\cellcolor[HTML]{C0C0C0}Evolvability issue}
        \\
        Code\_suggestion        &  0.378 & 1.459 & [1.378, 1.545] & ${<}0.0001$ \\
        has\_explanation       & -0.548 & 0.578 & [0.457, 0.731] & ${<}0.0001$ \\
        Conciseness            &  0.074 & 1.077 & [1.037, 1.118] & $0.0001$ \\
        Relevance              &  0.050 & 1.051 & [1.022, 1.080] & $0.0004$ \\
        Clarity                &  0.032 & 1.033 & [1.004, 1.062] & $0.0257$ \\
        exp\_Rule              &  0.097 & 1.102 & [1.027, 1.182] & $0.0067$ \\
        
        \bottomrule
      \end{tabular}
    }
  \par\smallskip
  \vspace{-0.5cm}
\end{table}

\section{Lessons Learned and Implications} \label{sec:implication}
\textbf{Revisiting Feedback Design: Actionability with Selective Explanations.} 
Our regression analysis suggests that the effectiveness of explanations depends on the nature of the problem being addressed. While the overall presence of explanations is negatively associated with usefulness, specific explanation styles, such as rules, examples, and subjective rationale, show positive effects, particularly for evolvability-related issues where clarity, conciseness, and relevance also emerge as significant predictors. In contrast, functional issues benefit more from brevity and direct, actionable fixes, with explanation playing a limited role. 
While prior work has highlighted the general importance of providing explanations in code review \cite{widyasari2025explaining}, it remains unclear which explanation styles are most effective for specific issue types. Our findings offer initial empirical evidence on this distinction, suggesting that agents should adapt their explanation strategy to the issue context.

\textbf{Context-Awareness as a Critical Factor for Trustworthy Agent-Generated Review.} 
A substantial portion of unresolved discussions falls under \textit{Incorrect Suggestion} and \textit{Intentional Design Decision} (Section~\ref{sec:RQ2-result}, indicating that agents often flag non-existent issues or fail to recognize project-specific choices. This aligns with A{\dh}alsteinsson et al. \cite{adhalsteinsson2025rethinking}, highlighting false positives as a key source of trust erosion. While modern LLMs offer larger context windows and agentic code exploration capabilities \cite{chen2025prometheus, robbes2026promises}, our findings show that broader context alone does not ensure contextually appropriate feedback. This underscores the need for agents to incorporate richer project-level signals, such as design rationale, conventions, and prior review history, to produce trustworthy feedback.

\textbf{Designing Agent-Generated Feedback for Diverse Developer Experience Levels.} Our analysis shows that core developers resolve most agent-generated comments, particularly those related to design and structural feedback, while peripheral developers are more involved in addressing functional defects (Section~\ref{sec:RQ2-result}). This aligns with prior findings that developer experience strongly influences code review outcomes~\cite{bosu2014impact, cynthia2026we}. In our study, we show that agents should adapt feedback accordingly by providing concrete suggestions for peripheral developers and higher-level insights for core developers, based on inferred experience.

\section{Related Work} 
Prior work has extensively studied what makes code review comments useful and how developers respond to them. Studies on human-generated reviews show that reviewer experience, proximity to changed code, and comment sentiment are key predictors of usefulness~\cite{bosu2015characteristics}, while approaches such as RevHelper~\cite{rahman2017predicting} and experience-aware automated review approaches~\cite{lin2024improving} aim to improve comment quality. More recently, researchers have examined the effectiveness of AI-generated review comments through resolution rates. For example, Sun et al. \cite{sun2025does} shows that hunk-level comments with inline suggestions are most effective, while Goldman et al. \cite{goldman2025types} identifies systematic differences across comment categories. Complementary studies further suggest that developers could treat AI suggestions as starting points rather than final solutions~\cite{ogenrwot2024patchtrack}, with only modest improvements in code quality~\cite{cihan2025automated}. 
Beyond comment characteristics, developer experience and project involvement significantly shape code review behavior and PR outcomes. Tsay et al. \cite{tsay2014influence} showed that contributors with stronger prior interaction with a project are more likely to have their PRs accepted, while Rahman et al. \cite{rahman2014insight} reported differences between successful and unsuccessful PRs based on project history and discussion context. At the review-process level, Thongtanunam et al. \cite{thongtanunam2017review} found that authors’ and reviewers’ prior involvement affects review participation, and Batoun et al. \cite{batoun2023empirical} showed that core and peripheral contributors exhibit distinct engagement patterns when responding to PRs.
With the rise of AI coding agents, recent work has begun to examine human–AI collaboration in software development, reporting increasing adoption of such tools \cite{shibu2025anthropic} but also highlighting challenges, including lower acceptance rates for complex tasks and issues with overly proactive or noisy feedback \cite{li2025rise, watanabe2025use, erlenhov2020empirical} and challenges around code quality \cite{takerngsaksiri2025human}. Nevertheless, prior work largely focuses on PR-level outcomes or developer perceptions, without examining how developers respond to individual agent-generated review comments.

In contrast, our work shifts the focus from PR-level decisions to comment-level resolution behavior of agent-generated feedback. By jointly analyzing comment characteristics and developer experience across a large-scale dataset of real-world interactions, we provide a comprehensive understanding of how and why developers act on agent-generated review comments, thereby bridging a critical gap between traditional code review research and emerging AI-assisted development.

\section{Threats to Validity} \label{sec:threats}
\textbf{\textit{External Validity.}} Our dataset consists of Python repositories on GitHub meeting activity thresholds (e.g., stars, PRs, and contributors), which may limit generalizability to other languages, small projects, or proprietary settings. We focused on Python-based repositories because many AI-agent and LLM ecosystems are built around Python, resulting in a large number of actively maintained repositories that employ AI-assisted development practices \cite{shah2026characterizing}. 
Agent identification based on login patterns may introduce misclassification, though manual checks mitigate this risk. 
Additionally, prompt configurations may vary across repositories and influence generated comments; however, such configurations are not publicly available and cannot be analyzed at scale.
Moreover, small sample sizes for Devin and Claude limit agent-specific conclusions.

\textbf{\textit{Internal Validity.}} 
We use \textit{Llama-3.1-70B} for comment classification and explanation annotation, selected after comparing three LLMs (Qwen-3-8B, Llama-3.1-70B, and GPT-4o) against human labels, where it achieved the strongest agreement ($\kappa = 0.74$, $J = 0.90$), though errors may remain.
Our logistic regression model shows low explanatory power (AUC = 0.58), suggesting that unobserved factors such as reviewer authority or project context may influence outcomes. 


\textbf{\textit{Construct Validity.}} We operationalize comment usefulness as resolution status, following prior work~\cite{goldman2025types}. However, resolution is an imperfect proxy, as comments may be resolved without being useful or remain unresolved despite being valuable. To mitigate this, we also consider comments as useful when developers explicitly acknowledge and apply the feedback. Nonetheless, some noise may remain in the usefulness label.
\section{Conclusion and Future Work} \label{sec:conclusion}
We presented the first empirical study on how developers respond to AI-agent-generated code review comments, analyzing $54{,}713$ comments from three coding agents across $341$ Python repositories on GitHub. Our results show that resolution rates vary across agents, with Copilot dominating both in volume and resolved comments, while Codex exhibits the lowest resolution rate. We further observe that core developers resolve most agent-generated feedback, particularly for design and evolvability-related comments, whereas peripheral developers are more involved in addressing functional defect comments. Through open card sorting of unresolved discussions, we identify ten recurring patterns, with \textit{Intentional Design Decision} and \textit{Incorrect Suggestion} being the most prevalent, highlighting agents’ limitations in capturing project-specific context. Finally, our regression analysis shows that usefulness is primarily driven by actionable feedback, with inline code suggestions as the strongest predictor, while selective explanation types provide complementary benefits depending on the issue context. 
Future work should explore how adaptive, context-aware agents can improve feedback effectiveness and developer trust. In particular, future studies can investigate agents that tailor their review comments based on the developer's role, prior contribution history and the specific project context. Such agents may provide more relevant explanations, avoid unnecessary or already-known suggestions for experienced developers and offer more actionable guidance to the peripheral contributors, leading to higher resolution rates, fewer unresolved discussions and greater trust in agent-generated feedback.


\balance
\small
\bibliographystyle{IEEEtran}
\bibliography{MANUSCRIPT}

@inproceedings{nguyen2025exploring,
  title={Exploring the Potential of Large Language Models in Fine-Grained Review Comment Classification},
  author={Nguyen, Linh and Liu, Chunhua and Lin, Hong Yi and Thongtanunam, Patanamon},
  booktitle={2025 IEEE International Conference on Source Code Analysis \& Manipulation (SCAM)},
  pages={43--54},
  year={2025},
  organization={IEEE}
}

@inproceedings{bouraffa2025not,
  title={Not One to Rule Them All: Mining Meaningful Code Review Orders From GitHub},
  author={Bouraffa, Abir and Brandt, Carolin and Zaidman, Andy and Maalej, Walid},
  booktitle={Proceedings of the 29th International Conference on Evaluation and Assessment in Software Engineering},
  pages={349--359},
  year={2025}
}

@inproceedings{dabic2021sampling,
  title={Sampling projects in github for MSR studies},
  author={Dabic, Ozren and Aghajani, Emad and Bavota, Gabriele},
  booktitle={2021 IEEE/ACM 18th International Conference on Mining Software Repositories (MSR)},
  pages={560--564},
  year={2021},
  organization={IEEE}
}

@inproceedings{kalliamvakou2014promises,
  title={The promises and perils of mining github},
  author={Kalliamvakou, Eirini and Gousios, Georgios and Blincoe, Kelly and Singer, Leif and German, Daniel M and Damian, Daniela},
  booktitle={Proceedings of the 11th working conference on mining software repositories},
  pages={92--101},
  year={2014}
}

@article{li2025rise,
  title={The rise of ai teammates in software engineering (se) 3.0: How autonomous coding agents are reshaping software engineering},
  author={Li, Hao and Zhang, Haoxiang and Hassan, Ahmed E},
  journal={arXiv preprint arXiv:2507.15003},
  year={2025}
}

@article{mantyla2008types,
  title={What types of defects are really discovered in code reviews?},
  author={M{\"a}ntyl{\"a}, Mika V and Lassenius, Casper},
  journal={IEEE Transactions on Software Engineering},
  volume={35},
  number={3},
  pages={430--448},
  year={2008},
  publisher={IEEE}
}

@article{turzo2024makes,
  title={What makes a code review useful to opendev developers? an empirical investigation},
  author={Turzo, Asif Kamal and Bosu, Amiangshu},
  journal={Empirical Software Engineering},
  volume={29},
  number={1},
  pages={6},
  year={2024},
  publisher={Springer}
}

@article{deng2023implicit,
  title={Implicit chain of thought reasoning via knowledge distillation},
  author={Deng, Yuntian and Prasad, Kiran and Fernandez, Roland and Smolensky, Paul and Chaudhary, Vishrav and Shieber, Stuart},
  journal={arXiv preprint arXiv:2311.01460},
  year={2023}
}

@inproceedings{sghaier2025harnessing,
  title={Harnessing Large Language Models for Curated Code Reviews},
  author={Sghaier, Oussama Ben and Weyssow, Martin and Sahraoui, Houari},
  booktitle={2025 IEEE/ACM 22nd International Conference on Mining Software Repositories (MSR)},
  pages={187--198},
  year={2025},
  organization={IEEE}
}

@article{mchugh2012interrater,
  title={Interrater reliability: the kappa statistic},
  author={McHugh, Mary L},
  journal={Biochemia medica},
  volume={22},
  number={3},
  pages={276--282},
  year={2012},
  publisher={Hrvatsko dru{\v{s}}tvo za medicinsku biokemiju i laboratorijsku medicinu}
}

@article{zheng2023judging,
  title={Judging llm-as-a-judge with mt-bench and chatbot arena},
  author={Zheng, Lianmin and Chiang, Wei-Lin and Sheng, Ying and Zhuang, Siyuan and Wu, Zhanghao and Zhuang, Yonghao and Lin, Zi and Li, Zhuohan and Li, Dacheng and Xing, Eric and others},
  journal={Advances in neural information processing systems},
  volume={36},
  pages={46595--46623},
  year={2023}
}

@inproceedings{goldman2025types,
  title={What Types of Code Review Comments Do Developers Most Frequently Resolve?},
  author={Goldman, Saul and Lin, Hong Yi and Pasuksmit, Jirat and Thongtanunam, Patanamon and Tantithamthavorn, Kla and Wang, Zhe and Zhang, Ray and Behnaz, Ali and Jiang, Fan and Siers, Michael and others},
  booktitle={2025 40th IEEE/ACM International Conference on Automated Software Engineering (ASE)},
  pages={3760--3765},
  year={2025},
  organization={IEEE}
}

@inproceedings{eyolfson2011time,
  title={Do time of day and developer experience affect commit bugginess?},
  author={Eyolfson, Jon and Tan, Lin and Lam, Patrick},
  booktitle={Proceedings of the 8th Working Conference on Mining Software Repositories},
  pages={153--162},
  year={2011}
}

@inproceedings{posnett2013dual,
  title={Dual ecological measures of focus in software development},
  author={Posnett, Daryl and D'Souza, Raissa and Devanbu, Premkumar and Filkov, Vladimir},
  booktitle={2013 35th International Conference on Software Engineering (ICSE)},
  pages={452--461},
  year={2013},
  organization={IEEE}
}

@inproceedings{mockus2010organizational,
  title={Organizational volatility and its effects on software defects},
  author={Mockus, Audris},
  booktitle={Proceedings of the eighteenth ACM SIGSOFT international symposium on Foundations of software engineering},
  pages={117--126},
  year={2010}
}

@inproceedings{robbes2013using,
  title={Using developer interaction data to compare expertise metrics},
  author={Robbes, Romain and R{\"o}thlisberger, David},
  booktitle={2013 10th Working Conference on Mining Software Repositories (MSR)},
  pages={297--300},
  year={2013},
  organization={IEEE}
}

@inproceedings{kononenko2016code,
  title={Code review quality: How developers see it},
  author={Kononenko, Oleksii and Baysal, Olga and Godfrey, Michael W},
  booktitle={Proceedings of the 38th international conference on software engineering},
  pages={1028--1038},
  year={2016}
}

@inproceedings{gousios2014exploratory,
  title={An exploratory study of the pull-based software development model},
  author={Gousios, Georgios and Pinzger, Martin and Deursen, Arie van},
  booktitle={Proceedings of the 36th international conference on software engineering},
  pages={345--355},
  year={2014}
}

@article{cynthia2026we,
  title={Are We All Using Agents the Same Way? An Empirical Study of Core and Peripheral Developers Use of Coding Agents},
  author={Cynthia, Shamse Tasnim and Das, Joy Krishan and Roy, Banani},
  journal={arXiv preprint arXiv:2601.20106},
  year={2026}
}

@article{mockus2002two,
  title={Two case studies of open source software development: Apache and Mozilla},
  author={Mockus, Audris and Fielding, Roy T and Herbsleb, James D},
  journal={ACM Transactions on Software Engineering and Methodology (TOSEM)},
  volume={11},
  number={3},
  pages={309--346},
  year={2002},
  publisher={ACM New York, NY, USA}
}

@inproceedings{joblin2017classifying,
  title={Classifying developers into core and peripheral: An empirical study on count and network metrics},
  author={Joblin, Mitchell and Apel, Sven and Hunsen, Claus and Mauerer, Wolfgang},
  booktitle={2017 IEEE/ACM 39th International Conference on Software Engineering (ICSE)},
  pages={164--174},
  year={2017},
  organization={IEEE}
}

@inproceedings{terceiro2010empirical,
  title={An empirical study on the structural complexity introduced by core and peripheral developers in free software projects},
  author={Terceiro, Antonio and Rios, Luiz Romario and Chavez, Christina},
  booktitle={2010 Brazilian Symposium on Software Engineering},
  pages={21--29},
  year={2010},
  organization={IEEE}
}

@book{spencer2009card,
  title={Card sorting: Designing usable categories},
  author={Spencer, Donna},
  year={2009},
  publisher={Rosenfeld Media}
}

@article{xu2020reinventing,
  title={Why reinventing the wheels? An empirical study on library reuse and re-implementation},
  author={Xu, Bowen and An, Le and Thung, Ferdian and Khomh, Foutse and Lo, David},
  journal={Empirical Software Engineering},
  volume={25},
  number={1},
  pages={755--789},
  year={2020},
  publisher={Springer}
}

@inproceedings{thongtanunam2022autotransform,
  title={Autotransform: Automated code transformation to support modern code review process},
  author={Thongtanunam, Patanamon and Pornprasit, Chanathip and Tantithamthavorn, Chakkrit},
  booktitle={Proceedings of the 44th international conference on software engineering},
  pages={237--248},
  year={2022}
}

@article{mcintosh2016empirical,
  title={An empirical study of the impact of modern code review practices on software quality},
  author={McIntosh, Shane and Kamei, Yasutaka and Adams, Bram and Hassan, Ahmed E},
  journal={Empirical Software Engineering},
  volume={21},
  number={5},
  pages={2146--2189},
  year={2016},
  publisher={Springer}
}

@inproceedings{morales2015code,
  title={Do code review practices impact design quality? a case study of the qt, vtk, and itk projects},
  author={Morales, Rodrigo and McIntosh, Shane and Khomh, Foutse},
  booktitle={2015 IEEE 22nd international conference on software analysis, evolution, and reengineering (SANER)},
  pages={171--180},
  year={2015},
  organization={IEEE}
}

@inproceedings{sadowski2018modern,
  title={Modern code review: a case study at google},
  author={Sadowski, Caitlin and S{\"o}derberg, Emma and Church, Luke and Sipko, Michal and Bacchelli, Alberto},
  booktitle={Proceedings of the 40th international conference on software engineering: Software engineering in practice},
  pages={181--190},
  year={2018}
}

@inproceedings{rigby2013convergent,
  title={Convergent contemporary software peer review practices},
  author={Rigby, Peter C and Bird, Christian},
  booktitle={Proceedings of the 2013 9th joint meeting on foundations of software engineering},
  pages={202--212},
  year={2013}
}

@inproceedings{adhalsteinsson2025rethinking,
  title={Rethinking code review workflows with llm assistance: An empirical study},
  author={A{\dh}alsteinsson, Fannar Steinn and Magn{\'u}sson, Bj{\"o}rn Borgar and Milicevic, Mislav and Davidsson, Adam Nirving and Cheng, Chih-Hong},
  booktitle={2025 ACM/IEEE International Symposium on Empirical Software Engineering and Measurement (ESEM)},
  pages={488--497},
  year={2025},
  organization={IEEE}
}

@inproceedings{frommgen2024resolving,
  title={Resolving code review comments with machine learning},
  author={Fr{\"o}mmgen, Alexander and Austin, Jacob and Choy, Peter and Ghelani, Nimesh and Kharatyan, Lera and Surita, Gabriela and Khrapko, Elena and Lamblin, Pascal and Manzagol, Pierre-Antoine and Revaj, Marcus and others},
  booktitle={Proceedings of the 46th international conference on software engineering: software engineering in practice},
  pages={204--215},
  year={2024}
}

@article{sun2025does,
  title={Does ai code review lead to code changes? a case study of github actions},
  author={Sun, Kexin and Kuang, Hongyu and Baltes, Sebastian and Zhou, Xin and Zhang, He and Ma, Xiaoxing and Rong, Guoping and Shao, Dong and Treude, Christoph},
  journal={arXiv preprint arXiv:2508.18771},
  year={2025}
}

@inproceedings{bosu2014impact,
  title={Impact of developer reputation on code review outcomes in oss projects: An empirical investigation},
  author={Bosu, Amiangshu and Carver, Jeffrey C},
  booktitle={Proceedings of the 8th ACM/IEEE international symposium on empirical software engineering and measurement},
  pages={1--10},
  year={2014}
}

@article{widyasari2025explaining,
  title={Explaining explanations: An empirical study of explanations in code reviews},
  author={Widyasari, Ratnadira and Zhang, Ting and Bouraffa, Abir and Maalej, Walid and Lo, David},
  journal={ACM Transactions on Software Engineering and Methodology},
  volume={34},
  number={6},
  pages={1--30},
  year={2025},
  publisher={ACM New York, NY}
}

@article{replication_package,
author = "Annonymous",
title = "{Replication Package}",
year = "2026",
month = "June",
url = "https://figshare.com/articles/conference_contribution/Replication_Package/32673702",
}

@inproceedings{tufano2022using,
  title={Using pre-trained models to boost code review automation},
  author={Tufano, Rosalia and Masiero, Simone and Mastropaolo, Antonio and Pascarella, Luca and Poshyvanyk, Denys and Bavota, Gabriele},
  booktitle={Proceedings of the 44th international conference on software engineering},
  pages={2291--2302},
  year={2022}
}

@inproceedings{decan2022use,
  title={On the use of github actions in software development repositories},
  author={Decan, Alexandre and Mens, Tom and Mazrae, Pooya Rostami and Golzadeh, Mehdi},
  booktitle={2022 IEEE International Conference on Software Maintenance and Evolution (ICSME)},
  pages={235--245},
  year={2022},
  organization={IEEE}
}

@article{da2025understanding,
  title={Understanding refactorings in Elixir functional language},
  author={da Matta Vegi, Lucas Francisco and Valente, Marco Tulio},
  journal={Empirical Software Engineering},
  volume={30},
  number={4},
  pages={108},
  year={2025},
  publisher={Springer}
}

@inproceedings{guzzi2013communication,
  title={Communication in open source software development mailing lists},
  author={Guzzi, Anja and Bacchelli, Alberto and Lanza, Michele and Pinzger, Martin and Van Deursen, Arie},
  booktitle={2013 10th Working Conference on Mining Software Repositories (MSR)},
  pages={277--286},
  year={2013},
  organization={IEEE}
}

@article{hirao2020code,
  title={Code reviews with divergent review scores: An empirical study of the openstack and qt communities},
  author={Hirao, Toshiki and McIntosh, Shane and Ihara, Akinori and Matsumoto, Kenichi},
  journal={IEEE Transactions on Software Engineering},
  volume={48},
  number={1},
  pages={69--81},
  year={2020},
  publisher={IEEE}
}

@inproceedings{bacchelli2013expectations,
  title={Expectations, outcomes, and challenges of modern code review},
  author={Bacchelli, Alberto and Bird, Christian},
  booktitle={2013 35th international conference on software engineering (ICSE)},
  pages={712--721},
  year={2013},
  organization={IEEE}
}

@misc{google_guide,
author = {Google},
title = {How to write code review comments | eng-practices},
howpublished = {\url{https://google.github.io/eng-practices/review/reviewer/comments.html}},
year = {2019}
}

@inproceedings{rahman2022example,
  title={Example Driven Code Review Explanation},
  author={Rahman, Shadikur and Koana, Umme Ayman and Nayebi, Maleknaz},
  booktitle={Proceedings of the 16th ACM/IEEE International Symposium on Empirical Software Engineering and Measurement},
  pages={307--312},
  year={2022}
}

@inproceedings{palvannan2023suggestion,
  title={Suggestion bot: analyzing the impact of automated suggested changes on code reviews},
  author={Palvannan, Nivishree and Brown, Chris},
  booktitle={2023 IEEE/ACM 5th International Workshop on Bots in Software Engineering (BotSE)},
  pages={33--37},
  year={2023},
  organization={IEEE}
}

@misc{chatgpt_codex_connector,
  author        ={OpenAI},
  title        = {Codex Code Review},
  year         = {2026},
  url          = {https://developers.openai.com/codex/integrations/github}
}

@inproceedings{bosu2015characteristics,
  title={Characteristics of useful code reviews: An empirical study at microsoft},
  author={Bosu, Amiangshu and Greiler, Michaela and Bird, Christian},
  booktitle={2015 IEEE/ACM 12th Working Conference on Mining Software Repositories},
  pages={146--156},
  year={2015},
  organization={IEEE}
}

@article{mchugh2013chi,
  title={The chi-square test of independence},
  author={McHugh, Mary L},
  journal={Biochemia medica},
  volume={23},
  number={2},
  pages={143--149},
  year={2013},
  publisher={Medicinska naklada}
}

@article{fay2010wilcoxon,
  title={Wilcoxon-Mann-Whitney or t-test? On assumptions for hypothesis tests and multiple interpretations of decision rules},
  author={Fay, Michael P and Proschan, Michael A},
  journal={Statistics surveys},
  volume={4},
  pages={1},
  year={2010}
}

@article{rani2023decade,
  title={A decade of code comment quality assessment: A systematic literature review},
  author={Rani, Pooja and Blasi, Arianna and Stulova, Nataliia and Panichella, Sebastiano and Gorla, Alessandra and Nierstrasz, Oscar},
  journal={Journal of Systems and Software},
  volume={195},
  pages={111515},
  year={2023},
  publisher={Elsevier}
}

@misc{github_graphql,
  author       = {{GitHub}},
  title        = {GitHub GraphQL API Documentation},
  year         = {2026},
  howpublished = {\url{https://docs.github.com/en/graphql}},
}

@misc{github_rest,
  author       = {{\ GitHub}},
  title        = {GitHub REST API Documentation},
  year         = {2026},
  howpublished = {\url{https://docs.github.com/en/rest}},
}

@misc{github_suggested_changes,
  author       = {{GitHub}},
  title        = {Incorporating feedback in your pull request},
  year         = {2026},
  howpublished = {\url{https://docs.github.com/en/pull-requests/collaborating-with-pull-requests/reviewing-changes-in-pull-requests/incorporating-feedback-in-your-pull-request}},
}

@inproceedings{rahman2017predicting,
  title={Predicting usefulness of code review comments using textual features and developer experience},
  author={Rahman, Mohammad Masudur and Roy, Chanchal K and Kula, Raula G},
  booktitle={2017 IEEE/ACM 14th International Conference on Mining Software Repositories (MSR)},
  pages={215--226},
  year={2017},
  organization={IEEE}
}

@article{ivchenko1998jaccard,
  title={On the jaccard similarity test},
  author={Ivchenko, GI and Honov, SA},
  journal={Journal of Mathematical Sciences},
  volume={88},
  number={6},
  pages={789--794},
  year={1998},
  publisher={Springer}
}

@misc{posthog2024pr42779,
  author       = {{PostHog}},
  title        = {Pull Request \#42779},
  year         = {2024},
  howpublished = {\url{https://github.com/posthog/posthog/pull/42779}}
}

@misc{opik_pr_3205,
  author       = {{comet-ml}},
  title        = {{Pull Request \#3205: Opik Repository}},
  year         = {2024},
  howpublished = {\url{https://github.com/comet-ml/opik/pull/3205}}
}

@misc{azure_pr_2854,
  author       = {{Azure-Samples}},
  title        = {{Pull Request \#2854: azure-search-openai-demo}},
  year         = {2024},
  howpublished = {\url{https://github.com/Azure-Samples/azure-search-openai-demo/pull/2854}}
}

@misc{sentry_pr_105249,
  author       = {{getsentry}},
  title        = {{Pull Request \#105249: sentry repository}},
  year         = {2024},
  howpublished = {\url{https://github.com/getsentry/sentry/pull/105249}}
}

@misc{ray_pr_54198,
  author       = {{ray-project}},
  title        = {{Pull Request \#54198: ray repo.}},
  year         = {2024},
  howpublished = {\url{https://github.com/ray-project/ray/pull/54198}}
}

@misc{ray_pr_56790,
  author       = {{ray\--project}},
  title        = {{Pull Request \#56790: ray repo.}},
  year         = {2025},
  howpublished = {\url{https://github.com/ray-project/ray/pull/56790}}
}

@misc{ray_pr_56752,
  author       = {{ray-project}},
  title        = {{Pull Request \#56752: ray repo.}},
  year         = {2025},
  howpublished = {\url{https://github.com/ray-project/ray/pull/56752}}
}

@misc{sentry_pr_104594,
  author       = {{getsentry}},
  title        = {{Pull Request \#104594: sentry repository}},
  year         = {2024},
  howpublished = {\url{https://github.com/getsentry/sentry/pull/104594}}
}

@misc{onyx_pr_6568,
  author       = {{onyx-dot-app}},
  title        = {{Pull Request \#6568: onyx repository}},
  year         = {2024},
  howpublished = {\url{https://github.com/onyx-dot-app/onyx/pull/6568}}
}

@misc{roboflow_pr_1812,
  author       = {{roboflow}},
  title        = {{Pull Request \#1812: inference repository}},
  year         = {2024},
  howpublished = {\url{https://github.com/roboflow/inference/pull/1812}}
}

@misc{litellm_pr_17152,
  author       = {{BerriAI}},
  title        = {{Pull Request \#17152: litellm repository}},
  year         = {2025},
  howpublished = {\url{https://github.com/BerriAI/litellm/pull/17152}}
}

@misc{kombu_pr_2333,
  author       = {{celery}},
  title        = {{Pull Request \#2333: kombu repository}},
  year         = {2024},
  howpublished = {\url{https://github.com/celery/kombu/pull/2333}}
}

@misc{fastdeploy_pr_5324,
  author       = {{paddlepaddle}},
  title        = {{Pull Request \#5324: fastdeploy repository}},
  year         = {2024},
  howpublished = {\url{https://github.com/paddlepaddle/fastdeploy/pull/5324}}
}

@misc{smolagents_pr_1448,
  author       = {{huggingface}},
  title        = {{Pull Request \#1448: smolagents repository}},
  year         = {2024},
  howpublished = {\url{https://github.com/huggingface/smolagents/pull/1448}}
}

@misc{openai_agents_pr_2173,
  author       = {{openai}},
  title        = {{Pull Request \#2173: openai-agents-python repository}},
  year         = {2025},
  howpublished = {\url{https://github.com/openai/openai-agents-python/pull/2173}}
}

@misc{sentry_pr_103695,
  author       = {{getsentry}},
  title        = {{Pull Request \#103695: sentry repository}},
  year         = {2024},
  howpublished = {\url{https://github.com/getsentry/sentry/pull/103695}}
}

@misc{mlflow_pr_19442,
  author       = {{mlflow}},
  title        = {{Pull Request \#19442: mlflow repository}},
  year         = {2025},
  howpublished = {\url{https://github.com/mlflow/mlflow/pull/19442}}
}

@misc{mlflow_pr_16382,
  author       = {{\ mlflow}},
  title        = {{Pull Request \#16382: mlflow repository}},
  year         = {2024},
  howpublished = {\url{https://github.com/mlflow/mlflow/pull/16382}}
}

@inproceedings{lin2024improving,
  title={Improving automated code reviews: Learning from experience},
  author={Lin, Hong Yi and Thongtanunam, Patanamon and Treude, Christoph and Charoenwet, Wachiraphan},
  booktitle={Proceedings of the 21st International Conference on Mining Software Repositories},
  pages={278--283},
  year={2024}
}

@inproceedings{tsay2014influence,
  title={Influence of social and technical factors for evaluating contribution in GitHub},
  author={Tsay, Jason and Dabbish, Laura and Herbsleb, James},
  booktitle={Proceedings of the 36th international conference on Software engineering},
  pages={356--366},
  year={2014}
}

@inproceedings{rahman2014insight,
  title={An insight into the pull requests of github},
  author={Rahman, Mohammad Masudur and Roy, Chanchal K},
  booktitle={Proceedings of the 11th working conference on mining software repositories},
  pages={364--367},
  year={2014}
}

@article{thongtanunam2017review,
  title={Review participation in modern code review: An empirical study of the android, Qt, and OpenStack projects},
  author={Thongtanunam, Patanamon and McIntosh, Shane and Hassan, Ahmed E and Iida, Hajimu},
  journal={Empirical Software Engineering},
  volume={22},
  number={2},
  pages={768--817},
  year={2017},
  publisher={Springer}
}

@article{batoun2023empirical,
  title={An empirical study on GitHub pull requests’ reactions},
  author={Batoun, Mohamed Amine and Yung, Ka Lai and Tian, Yuan and Sayagh, Mohammed},
  journal={ACM Transactions on Software Engineering and Methodology},
  volume={32},
  number={6},
  pages={1--35},
  year={2023},
  publisher={ACM New York, NY}
}

@misc{shibu2025anthropic,
  author       = {Shibu, Sherin},
  title        = {{`I do have a fair amount of concern.': The CEO of \$61 billion Anthropic says AI will take over a crucial part of software engineers’ jobs within a year}},
  year         = {2025},
  howpublished = {\url{https://www.entrepreneur.com/business-news/anthropic-ceo-predicts-ai-will-take-over-coding-in-12-months/488533}}
}

@article{watanabe2025use,
  title={On the Use of Agentic Coding: An Empirical Study of Pull Requests on GitHub},
  author={Watanabe, Miku and Li, Hao and Kashiwa, Yutaro and Reid, Brittany and Iida, Hajimu and Hassan, Ahmed E},
  journal={arXiv preprint arXiv:2509.14745},
  year={2025}
}

@inproceedings{erlenhov2020empirical,
  title={An empirical study of bots in software development: Characteristics and challenges from a practitioner’s perspective},
  author={Erlenhov, Linda and Neto, Francisco Gomes De Oliveira and Leitner, Philipp},
  booktitle={Proceedings of the 28th ACM joint meeting on european software engineering conference and symposium on the foundations of software engineering},
  pages={445--455},
  year={2020}
}

@inproceedings{cihan2025automated,
  title={Automated code review in practice},
  author={Cihan, Umut and Haratian, Vahid and {\.I}{\c{c}}{\"o}z, Arda and G{\"u}l, Mert Kaan and Devran, {\"O}mercan and Bayendur, Emircan Furkan and U{\c{c}}ar, Baykal Mehmet and T{\"u}z{\"u}n, Eray},
  booktitle={2025 IEEE/ACM 47th International Conference on Software Engineering: Software Engineering in Practice (ICSE-SEIP)},
  pages={425--436},
  year={2025},
  organization={IEEE}
}

@inproceedings{ogenrwot2024patchtrack,
  title={PatchTrack: Analyzing ChatGPT's Impact on Software Patch Decision-Making in Pull Requests},
  author={Ogenrwot, Daniel and Businge, John},
  booktitle={Proceedings of the 39th IEEE/ACM International Conference on Automated Software Engineering},
  pages={2480--2481},
  year={2024}
}

@inproceedings{takerngsaksiri2025human,
  title={Human-in-the-loop software development agents},
  author={Takerngsaksiri, Wannita and Pasuksmit, Jirat and Thongtanunam, Patanamon and Tantithamthavorn, Chakkrit and Zhang, Ruixiong and Jiang, Fan and Li, Jing and Cook, Evan and Chen, Kun and Wu, Ming},
  booktitle={2025 IEEE/ACM 47th International Conference on Software Engineering: Software Engineering in Practice (ICSE-SEIP)},
  pages={342--352},
  year={2025},
  organization={IEEE}
}

@article{chen2025prometheus,
  title={Prometheus: Unified knowledge graphs for issue resolution in multilingual codebases},
  author={Chen, Zimin and Pan, Yue and Lu, Siyu and Xu, Jiayi and Goues, Claire Le and Monperrus, Martin and Ye, He},
  journal={arXiv preprint arXiv:2507.19942},
  year={2025}
}

@article{robbes2026promises,
  title={Promises, Perils, and (Timely) Heuristics for Mining Coding Agent Activity},
  author={Robbes, Romain and Matricon, Th{\'e}o and Degueule, Thomas and Hora, Andre and Zacchiroli, Stefano},
  journal={arXiv preprint arXiv:2601.18345},
  year={2026}
}

@misc{compiler_explorer_pr_1740,
  author       = {{Compiler Explorer Contributors}},
  title        = {{Pull Request \#1740: infra repository}},
  year         = {2024},
  howpublished = {\url{https://github.com/compiler-explorer/infra/pull/1740}}
}

@misc{langsmith_sdk_pr_2030,
  author       = {{LangChain Contributors}},
  title        = {{Pull Request \#2030: langsmith-sdk repository}},
  year         = {2024},
  howpublished = {\url{https://github.com/langchain-ai/langsmith-sdk/pull/2030}}
}

@article{lin2025leveraging,
author = {Lin, Hong Yi and Thongtanunam, Patanamon and Treude, Christoph and Godfrey, Michael W. and Liu, Chunhua and Charoenwet, Wachiraphan},
title = {Leveraging Reviewer Experience in Code Review Comment Generation},
journal={ACM Transactions on Software Engineering and Methodology (TOSEM)},
year = {2025},
publisher = {Association for Computing Machinery},
address = {New York, NY, USA},
issn = {1049-331X},
note = {Just Accepted},
}

@article{hasan2026empirical,
  title={An empirical study of testing practices in open source AI agent frameworks and agentic applications},
  author={Hasan, Mohammed Mehedi and Li, Hao and Fallahzadeh, Emad and Rajbahadur, Gopi Krishnan and Adams, Bram and Hassan, Ahmed E},
  journal={Empirical Software Engineering},
  volume={31},
  number={5},
  pages={124},
  year={2026},
  publisher={Springer}
}

@article{openja2024empirical,
  title={An empirical study of testing machine learning in the wild},
  author={Openja, Moses and Khomh, Foutse and Foundjem, Armstrong and Jiang, Zhen Ming and Abidi, Mouna and Hassan, Ahmed E},
  journal={ACM transactions on software engineering and methodology},
  volume={34},
  number={1},
  pages={1--63},
  year={2024},
  publisher={ACM New York, NY}
}

@article{shah2026characterizing,
  title={Characterizing faults in agentic AI: A taxonomy of types, symptoms, and root causes},
  author={Shah, Mehil B and Morovati, Mohammad Mehdi and Rahman, Mohammad Masudur and Khomh, Foutse},
  journal={arXiv preprint arXiv:2603.06847},
  year={2026}
}

\end{document}